\documentclass[aps,prd,twocolumn,amsmath,amssymb,superscriptaddress,floatfix,preprintnumbers]{revtex4-1}
\usepackage{graphicx}
\usepackage[space]{grffile}
\usepackage{dcolumn}
\usepackage{bm}
\usepackage{color}
\usepackage{siunitx}
\usepackage{multirow}
\usepackage{xspace}
\usepackage{url}
\usepackage{lineno}
\usepackage[english]{babel}
\usepackage{natbib}
\usepackage{csvsimple}
\usepackage{enumitem}
\usepackage{float}
\usepackage{placeins}
\usepackage{verbatim}
\usepackage{hyperref}   
\usepackage{silence}    
\usepackage{longtable}
\usepackage{etoolbox}
\usepackage[normalem]{ulem}
\usepackage{tcolorbox}
\usepackage{longtable}
\usepackage{lineno}
\usepackage{afterpage}

\newcommand{\numu}{\ensuremath{\nu_\mu}\xspace}

\newcommand{\numucc}{\ensuremath{\nu_{\mu}~\text{CC}}\xspace}
\newcommand{\tmu}{\ensuremath{T_\mu}\xspace}
\newcommand{\cosmu}{\ensuremath{\cos\theta_\mu}\xspace}
\newcommand{\eav}{\ensuremath{E_\mathrm{avail}}\xspace}

\newcommand{\dd}{\ensuremath{\dfrac{\mathrm{d}^2\sigma}{\mathrm{d}\cos\theta_\mu \mathrm{d} T_\mu}}\xspace}

\newcommand{\sd}{\ensuremath{\dfrac{\mathrm{d}\sigma}{\mathrm{d} Q^2}}\xspace}

\newcommand{\deltacp}{\ensuremath{\delta_\mathrm{CP}}\xspace}

\newcommand{\valencia}{Valencia\xspace}
\newcommand{\susa}{SuSAv2\xspace}

\newcommand{\nova}{NOvA\xspace}

\newcommand{\Enu}{\ensuremath{E_\nu}\xspace}
\newcommand{\Qsq}{\ensuremath{Q^2}\xspace}

\begin{document}

\title{Measurement of the double-differential cross section of muon-neutrino charged-current interactions with low hadronic energy in the NOvA Near Detector} 

\newcommand{\ANL}{Argonne National Laboratory, Argonne, Illinois 60439, 
USA}
\newcommand{\Bandirma}{Bandirma Onyedi Eyl\"ul University, Faculty of 
Engineering and Natural Sciences, Engineering Sciences Department, 
10200, Bandirma, Balıkesir, Turkey}
\newcommand{\ICS}{Institute of Computer Science, The Czech 
Academy of Sciences, 
182 07 Prague, Czech Republic}
\newcommand{\IOP}{Institute of Physics, The Czech 
Academy of Sciences, 
182 21 Prague, Czech Republic}
\newcommand{\Atlantico}{Universidad del Atlantico,
Carrera 30 No.\ 8-49, Puerto Colombia, Atlantico, Colombia}
\newcommand{\BHU}{Department of Physics, Institute of Science, Banaras 
Hindu University, Varanasi, 221 005, India}
\newcommand{\UCLA}{Physics and Astronomy Department, UCLA, Box 951547, Los 
Angeles, California 90095-1547, USA}
\newcommand{\Caltech}{California Institute of 
Technology, Pasadena, California 91125, USA}
\newcommand{\Cochin}{Department of Physics, Cochin University
of Science and Technology, Kochi 682 022, India}
\newcommand{\Charles}
{Charles University, Faculty of Mathematics and Physics,
 Institute of Particle and Nuclear Physics, Prague, Czech Republic}
\newcommand{\Cincinnati}{Department of Physics, University of Cincinnati, 
Cincinnati, Ohio 45221, USA}
\newcommand{\CSU}{Department of Physics, Colorado 
State University, Fort Collins, CO 80523-1875, USA}
\newcommand{\CTU}{Czech Technical University in Prague,
Brehova 7, 115 19 Prague 1, Czech Republic}
\newcommand{\Dallas}{Physics Department, University of Texas at Dallas,
800 W. Campbell Rd. Richardson, Texas 75083-0688, USA}
\newcommand{\DallasU}{University of Dallas, 1845 E 
Northgate Drive, Irving, Texas 75062 USA}
\newcommand{\Delhi}{Department of Physics and Astrophysics, University of 
Delhi, Delhi 110007, India}
\newcommand{\JINR}{Joint Institute for Nuclear Research,  
Dubna, Moscow region 141980, Russia}
\newcommand{\Erciyes}{
Department of Physics, Erciyes University, Kayseri 38030, Turkey}
\newcommand{\FNAL}{Fermi National Accelerator Laboratory, Batavia, 
Illinois 60510, USA}
\newcommand{\FSU}{Florida State University, Tallahassee, Florida 32306, USA}
\newcommand{\UFG}{Instituto de F\'{i}sica, Universidade Federal de 
Goi\'{a}s, Goi\^{a}nia, Goi\'{a}s, 74690-900, Brazil}
\newcommand{\Guwahati}{Department of Physics, IIT Guwahati, Guwahati, 781 
039, India}
\newcommand{\Harvard}{Department of Physics, Harvard University, 
Cambridge, Massachusetts 02138, USA}
\newcommand{\Houston}{Department of Physics, 
University of Houston, Houston, Texas 77204, USA}
\newcommand{\IHyderabad}{Department of Physics, IIT Hyderabad, Hyderabad, 
502 205, India}
\newcommand{\Hyderabad}{School of Physics, University of Hyderabad, 
Hyderabad, 500 046, India}
\newcommand{\IIT}{Illinois Institute of Technology,
Chicago IL 60616, USA}
\newcommand{\Indiana}{Indiana University, Bloomington, Indiana 47405, 
USA}
\newcommand{\INR}{Institute for Nuclear Research of Russia, Academy of 
Sciences 7a, 60th October Anniversary prospect, Moscow 117312, Russia}
\newcommand{\Iowa}{Department of Physics and Astronomy, Iowa State 
University, Ames, Iowa 50011, USA}
\newcommand{\Irvine}{Department of Physics and Astronomy, 
University of California at Irvine, Irvine, California 92697, USA}
\newcommand{\Jammu}{Department of Physics and Electronics, University of 
Jammu, Jammu Tawi, 180 006, Jammu and Kashmir, India}
\newcommand{\Lebedev}{Nuclear Physics and Astrophysics Division, Lebedev 
Physical 
Institute, Leninsky Prospect 53, 119991 Moscow, Russia}
\newcommand{\Magdalena}{Universidad del Magdalena, Carrera 32 No 22-08 Santa Marta, Colombia}
\newcommand{\MSU}{Department of Physics and Astronomy, Michigan State 
University, East Lansing, Michigan 48824, USA}
\newcommand{\Crookston}{Math, Science and Technology Department, University 
of Minnesota Crookston, Crookston, Minnesota 56716, USA}
\newcommand{\Duluth}{Department of Physics and Astronomy, 
University of Minnesota Duluth, Duluth, Minnesota 55812, USA}
\newcommand{\Minnesota}{School of Physics and Astronomy, University of 
Minnesota Twin Cities, Minneapolis, Minnesota 55455, USA}
\newcommand{\Mississippi}{University of Mississippi, University, Mississippi 38677, USA}
\newcommand{\NISER}{National Institute of Science Education and Research,
Khurda, 752050, Odisha, India}
\newcommand{\Oxford}{Subdepartment of Particle Physics, 
University of Oxford, Oxford OX1 3RH, United Kingdom}
\newcommand{\Panjab}{Department of Physics, Panjab University, 
Chandigarh, 160 014, India}
\newcommand{\Pitt}{Department of Physics, 
University of Pittsburgh, Pittsburgh, Pennsylvania 15260, USA}
\newcommand{\QMU}{Particle Physics Research Centre, 
Department of Physics and Astronomy,
Queen Mary University of London,
London E1 4NS, United Kingdom}
\newcommand{\RAL}{Rutherford Appleton Laboratory, Science 
and 
Technology Facilities Council, Didcot, OX11 0QX, United Kingdom}
\newcommand{\SAlabama}{Department of Physics, University of 
South Alabama, Mobile, Alabama 36688, USA} 
\newcommand{\Carolina}{Department of Physics and Astronomy, University of 
South Carolina, Columbia, South Carolina 29208, USA}
\newcommand{\SDakota}{South Dakota School of Mines and Technology, Rapid 
City, South Dakota 57701, USA}
\newcommand{\SMU}{Department of Physics, Southern Methodist University, 
Dallas, Texas 75275, USA}
\newcommand{\Stanford}{Department of Physics, Stanford University, 
Stanford, California 94305, USA}
\newcommand{\Sussex}{Department of Physics and Astronomy, University of 
Sussex, Falmer, Brighton BN1 9QH, United Kingdom}
\newcommand{\Syracuse}{Department of Physics, Syracuse University,
Syracuse NY 13210, USA}
\newcommand{\Tennessee}{Department of Physics and Astronomy, 
University of Tennessee, Knoxville, Tennessee 37996, USA}
\newcommand{\Texas}{Department of Physics, University of Texas at Austin, 
Austin, Texas 78712, USA}
\newcommand{\Tufts}{Department of Physics and Astronomy, Tufts University, Medford, 
Massachusetts 02155, USA}
\newcommand{\UCL}{Physics and Astronomy Department, University College 
London, 
Gower Street, London WC1E 6BT, United Kingdom}
\newcommand{\Virginia}{Department of Physics, University of Virginia, 
Charlottesville, Virginia 22904, USA}
\newcommand{\WSU}{Department of Mathematics, Statistics, and Physics,
 Wichita State University, 
Wichita, Kansas 67260, USA}
\newcommand{\WandM}{Department of Physics, William \& Mary, 
Williamsburg, Virginia 23187, USA}
\newcommand{\Wisconsin}{Department of Physics, University of 
Wisconsin-Madison, Madison, Wisconsin 53706, USA}
\newcommand{\deceased}{Deceased.}
\affiliation{\ANL}
\affiliation{\Atlantico}
\affiliation{\Bandirma}
\affiliation{\BHU}
\affiliation{\Caltech}
\affiliation{\Charles}
\affiliation{\Cincinnati}
\affiliation{\Cochin}
\affiliation{\CSU}
\affiliation{\CTU}
\affiliation{\Delhi}
\affiliation{\Erciyes}
\affiliation{\FNAL}
\affiliation{\FSU}
\affiliation{\UFG}
\affiliation{\Guwahati}
\affiliation{\Houston}
\affiliation{\Hyderabad}
\affiliation{\IHyderabad}
\affiliation{\IIT}
\affiliation{\Indiana}
\affiliation{\ICS}
\affiliation{\INR}
\affiliation{\IOP}
\affiliation{\Iowa}
\affiliation{\Irvine}
\affiliation{\JINR}
\affiliation{\Magdalena}
\affiliation{\MSU}
\affiliation{\Duluth}
\affiliation{\Minnesota}
\affiliation{\Mississippi}
\affiliation{\NISER}
\affiliation{\Panjab}
\affiliation{\Pitt}
\affiliation{\QMU}
\affiliation{\SAlabama}
\affiliation{\Carolina}
\affiliation{\SMU}
\affiliation{\Sussex}
\affiliation{\Syracuse}
\affiliation{\Texas}
\affiliation{\Tufts}
\affiliation{\UCL}
\affiliation{\Virginia}
\affiliation{\WSU}
\affiliation{\WandM}
\affiliation{\Wisconsin}

\author{M.~A.~Acero}
\affiliation{\Atlantico}

\author{B.~Acharya}
\affiliation{\Mississippi}

\author{P.~Adamson}
\affiliation{\FNAL}



\author{L.~Aliaga}
\affiliation{\FNAL}






\author{N.~Anfimov}
\affiliation{\JINR}


\author{A.~Antoshkin}
\affiliation{\JINR}


\author{E.~Arrieta-Diaz}
\affiliation{\Magdalena}

\author{L.~Asquith}
\affiliation{\Sussex}


\author{A.~Aurisano}
\affiliation{\Cincinnati}


\author{A.~Back}
\affiliation{\Indiana}
\affiliation{\Iowa}



% ...

\collaboration{The NOvA Collaboration}
\noaffiliation




\preprint{FERMILAB-PUB-24-0654-PPD}

\begin{abstract}
The \nova collaboration reports cross-section measurements for \numu charged-current interactions with low hadronic energy (maximum kinetic energy of 250 MeV for protons and 175 MeV for pions) in the NOvA Near Detector. The results are presented as a double-differential cross section as a function of the direct observables of the final-state muon kinematics. Results are also presented as a single-differential cross section as a function of the derived square of the four-momentum transfer, \Qsq, and as a function of the derived neutrino energy. The data correspond to an accumulated 8.09$\times10^{20}$ protons-on-target (POT) in the neutrino mode of the NuMI beam, with a narrow band of neutrino energies peaked at \SI{1.8}{GeV}. The analysis provides a sample of neutrino--nucleus interactions with an enhanced fraction of quasi-elastic and two-particle-two-hole (2p2h) interactions. This enhancement allows quantitative comparisons with various nuclear models. We find strong disagreement between data and theory-based models in various regions of the muon kinematic phase space, especially in the forward muon direction.
\end{abstract}

\maketitle

\section{Introduction}
\label{sec:introduction}
Current and future neutrino oscillation experiments aim to substantially improve the measurements of the PMNS oscillation parameters~\cite{bib:osctheory1,bib:osctheory2} and determine the neutrino mass ordering. Experiments such as NOvA~\cite{bib:NOvAOsc2021} and T2K~\cite{bib:T2KoscMarch2023} are sensitive to the neutrino mass ordering and to the parameter \deltacp, which describes the extent to which neutrinos violate charge-parity symmetry. Future experiments, such as DUNE~\cite{bib:DUNETDR} and HyperK~\cite{bib:HyperK}, have been designed to make precise measurements of many of the neutrino oscillation parameters and determine \deltacp. To maximize the potential of these measurements, improvements in neutrino--nucleus cross-section modeling are needed as they are expected to be a leading systematic uncertainty in these future experiments~\cite{bib:DUNETDR,bib:DUNENDCDR,bib:HyperK}. 

The measurement of neutrino oscillation parameters requires relating the observed kinematics of the final-state particles from neutrino--nucleus interactions in the detector to the incoming neutrino energy, a process that depends on neutrino interaction models~\cite{bib:NuSTECWP}.  Furthermore, the event selection criteria and any needed corrections, such as the detector acceptance, the relationship between the desired signal and backgrounds, and the detector resolution, also depend on neutrino interaction models.  Direct measurements of these cross sections are, therefore, invaluable for neutrino oscillation experiments.

One of the challenges in the field of neutrino--nucleus scattering is the modeling of nuclear effects and their impact on the final state particle kinematics~\cite{PhysRevD.105.092004,Katori_2018}. For example, neutrino interactions with a correlated pair of nucleons may result in two-particle-two-hole (2p2h) processes. The dominant 2p2h process occurs when a virtual meson is exchanged between the two correlated nucleons in the scattering process.  These meson exchange currents (MEC) are significant when heavier nuclear targets are involved.  

Recent results from MINERvA~\cite{bib:MINERvAMELowq3}, MicroBooNE~\cite{bib:uBooNE2023CCQElike}, NOvA~\cite{bib:NOvANuMuCCInc}, and T2K~\cite{bib:T2K2023CC0pi} show discrepancies with scattering models implemented in various neutrino event generators. There are also apparent discrepancies across the experiments’ datasets. For instance, at lower beam energies (sub-GeV), there is a disagreement between recent MicroBooNE 2p2h-enhanced measurements and the MicroBooNE tuned model that includes a tuning to the T2K data~\cite{bib:uBooNE2023CC2parxiv}. At higher beam energies (few-GeV), as shown below, the predicted 2p2h distributions using the MINERvA- and NOvA-tuned models do not agree.   
In both cases, there are other nuclear effects that are poorly constrained, such as long-range correlations~\cite{bib:Martini_2009,bib:Nieves_2004,bib:Pandey_2015}, pion absorption~\cite{bib:Oset,bib:NuWro1,bib:GENIE1,bib:GiBUU1,bib:NEUT}, and low \Qsq resonance behavior~\cite{bib:miniboone-res, bib:minos-qe, bib:minerva-pi-1, bib:minerva-pi-2, bib:t2k1pi}. These impact the interpretation of neutrino--nucleus interactions in measurements.

The NOvA collaboration has previously published a measurement of the inclusive muon-neutrino charged-current (CC) double-differential cross section as a function of muon kinematics~\cite{bib:NOvANuMuCCInc}. In this paper, we build on that analysis with a similar selection but with tighter constraints to report the double-differential cross section for interactions with low hadronic energy. This sample includes phase-space regions sensitive to the 2p2h process. We take advantage of this feature to compare our results more incisively with various 2p2h models. 
The low hadronic energy sample is defined as neutrino interactions which result in no protons with kinetic energy above \SI{250}{MeV} and no pions with kinetic energy above \SI{175}{MeV}. 
We report flux-integrated double-differential cross-section measurements of \numu CC interactions with low hadronic energy in the NOvA Near Detector with respect to the outgoing muon kinetic energy and cosine of the muon angle with respect to the beam direction. 
 Additionally, we report single-differential cross sections with respect to the neutrino energy (\Enu) and the square of the four-momentum transfer (\Qsq). The results presented correspond to 8.09$\times10^{20}$ protons delivered to the NuMI production target (POT) between November 2014 and February 2017.

\section{Beam and Detectors}
\label{sec:experiment}

NOvA is primarily a long-baseline neutrino oscillation experiment~\cite{bib:NOvAOsc2021}. It consists of two functionally identical detectors: the Near Detector (ND) and the Far Detector (FD). The ND is located \SI{100}{\meter} underground at Fermilab, Batavia, IL, approximately \SI{1}{\kilo\meter} from the neutrino production target. The FD is located on the surface near Ash River, MN, \SI{810}{km} from the target. 

The NuMI complex at Fermilab provides the neutrino beam for NOvA~\cite{bib:NuMI2016}. The neutrino beam production starts with \SI{120}{GeV} protons from the Fermilab Main Injector striking a \SI{1.2}{m}-long graphite target producing a hadronic cascade. Charged hadrons escaping the target are focused by two magnetic horns. Pions and kaons decay to neutrinos in a \SI{675}{m}-long volume filled with helium. The polarity of the horns can be tuned to focus either positive or negative particles, resulting a beam enriched in either neutrinos or antineutrinos. Both NOvA detectors are situated \SI{14.6}{mrad} off-axis from the NuMI beam, resulting in a flux of neutrinos at the ND peaked at \SI{1.8}{GeV} with a width of about \SI{0.7}{GeV} with a long tail extending beyond \SI{5}{GeV}. For this analysis, we use the neutrino-mode beam which is composed of 97.5\% muon neutrinos, 1.8\% muon antineutrinos, and 0.7\% electron neutrinos and antineutrinos for neutrino energies between 1 and 5 GeV. 

The NOvA ND is a segmented tracking calorimeter with square planes assembled from rounded rectangular PVC cells whose long axis is transverse to the neutrino beam.
The cells are \SI{3.9}{\cm} wide, \SI{6.6}{\cm} deep (in the direction of the beam), and \SI{3.9}{\meter} long. Planes of cells are arranged in alternating horizontal and vertical directions for three-dimensional reconstruction. The liquid scintillator is a mix of 95\% mineral oil and 5\% pseudocumene with trace concentrations of wavelength shifting fluors.  The chemical composition of the fiducial volume is 67\% carbon, 16\% chlorine, 11\% hydrogen, 3\% titanium and 3\% oxygen. The fully active region of the detector, consisting entirely of PVC extrusions filled with liquid scintillator, is \SI{12.7}{\meter} long. Downstream of the fully active detector is the muon catcher, constructed of pair of planes of PVC cells separated by slabs of steel, designed to range out and measure muons with energies up to \SI{2.5}{GeV}.  The muon catcher is \SI{3}{\meter} deep, \SI{3.9}{\meter} wide and \SI{2.6}{\meter} high and covers the bottom two-thirds of the detector. Each PVC cell has a wavelength-shifting (WLS) fiber to collect and direct the light to an avalanche photodiode for digitization by custom front-end electronics. All data associated with a NuMI trigger and above a noise-vetoing threshold are stored for further processing. 

\section{Simulation and Reconstruction}
\label{sec:simulation}

The neutrino flux prediction is based on a detailed simulation of the NuMI beamline using \textsc{Geant4 v9.2.p03} with the FTFP\_BERT hadronic physics list~\cite{bib:GEANT4}. The flux model is constrained using the PPFX package (Package to Predict the FluX)~\cite{bib:MINERvAPPFX}. PPFX uses data from proton--carbon interactions and other thin-target hadron production data ~\cite{bib:Paley-MIPP, bib:Alt-NA49, bib:Abgrall-NA61, bib:Barton-83, bib:Seun-07, bib:Lebedev-07, bib:Tinti-10, bib:Baatar-NA49, bib:Skubic-78, bib:Denisov-73, bib:Carroll-79, bib:Abe-13-13, bib:Cronin-57, bib:Allaby-69, bib:Longo-62, bib:Bobchenko-79, bib:Fedorov-78, bib:Abrams-70} to calculate the predictions and propagate uncertainties. 

Neutrino interactions for this analysis are simulated using the \textsc{GENIE v2.12.2} event generator~\cite{bib:GENIE1,bib:GENIE2}. The Llewellyn Smith formalism~\cite{bib:LlewellynSmith} with axial mass, $M_A = $ \SI{0.99}{GeV/c^2},
is used for quasi-elastic (QE) interactions. The nuclear environment for QE interactions is simulated with the Relativistic Fermi Gas (RFG) model~\cite{bib:SmithMoniz}, using the Bodek--Ritchie high-momentum tail on the nucleon momentum distribution to account for short-range correlations. The Rein--Sehgal model~\cite{bib:ResReinSeghal} is used to simulate resonant (Res) production. The Deep Inelastic Scattering (DIS) model is based on the Bodek--Yang prescription~\cite{bib:BodekYang}. We model 2p2h processes using the Empirical 2p2h model~\cite{bib:MECModels}. Final-state interactions (FSI) are simulated by the hN semiclassical intranuclear cascade model in which pion interaction probabilities are assigned according to Oset et al.~\cite{bib:GENIEFSI} and pion--nucleon scattering data.

NOvA applies corrections to the default GENIE version based on measurements and improved models. These corrections include setting $M_A=$\SI{1.04}{GeV/c^{2}} for the CC QE cross section, reducing non-resonant single pion production by 57\%, applying weights to reproduce the \valencia group's Random Phase Approximation (RPA) calculation of the nuclear field for QE, as well as weights to sculpt the Res cross section to have a similar dependence on $Q^2$ as the QE~\cite{bib:MINERvAtuning} and increasing the predicted rate of DIS events with hadronic mass W $>$ \SI{1.7}{GeV} by 10\%. Finally, the rate of Empirical 2p2h interactions is adjusted to match a subset of \numucc NOvA ND data, changing the shape of the underlying model. The resulting predictions after these corrections is referred to as the NOvA tune v1~\cite{bib:NOvATune2020}. 

Neutrino interactions are simulated in a detailed description of the NOvA ND geometry and the surrounding rock. Time-dependent variations of the proton beam intensity and active detector channels are also simulated to model actual data-taking conditions. The energy deposited in the propagation of the particles is simulated using \textsc{Geant4 v10.1.p03}. The scintillator response and the fiber attenuation are modeled using NOvA measurements~\cite{bib:NOvAsimulation}. The Birks' suppression of the light yield and the electronic readout response are tuned to NOvA test-stand measurements~\cite{bib:JINRTestStand}.

The event reconstruction of the \numucc interactions is identical to that used in \cite{bib:NOvANuMuCCInc}.  Three-dimensional trajectories of charged particles (tracks) are formed via a Kalman filter algorithm~\cite{bib:Kalman} that uses energy depositions (hits) in cells that are correlated in space and time windows~\cite{bib:BairdThesis}.  The upstream end of the muon track is considered the interaction vertex point.  

\section{Event Selection and Signal Definition}
\label{sec:eventsample}

The muon identification algorithm (\texttt{MuonID}), identical to that used in \cite{bib:NOvANuMuCCInc}, uses $dE/dx$ and scattering log-likelihood differences between muons and pions and the average $dE/dx$ in the cells of the reconstruct track trajectory's last 10 and \SI{40}{cm} to identify muons in the final state. Except for the muon candidate, all tracks and other energy deposits associated with the event are required to be contained in a sub-volume of the active region; muon tracks, however, are allowed to enter the muon catcher. 
\begin{figure}[!htbp]
    \centering
    \includegraphics[width=0.45\textwidth]{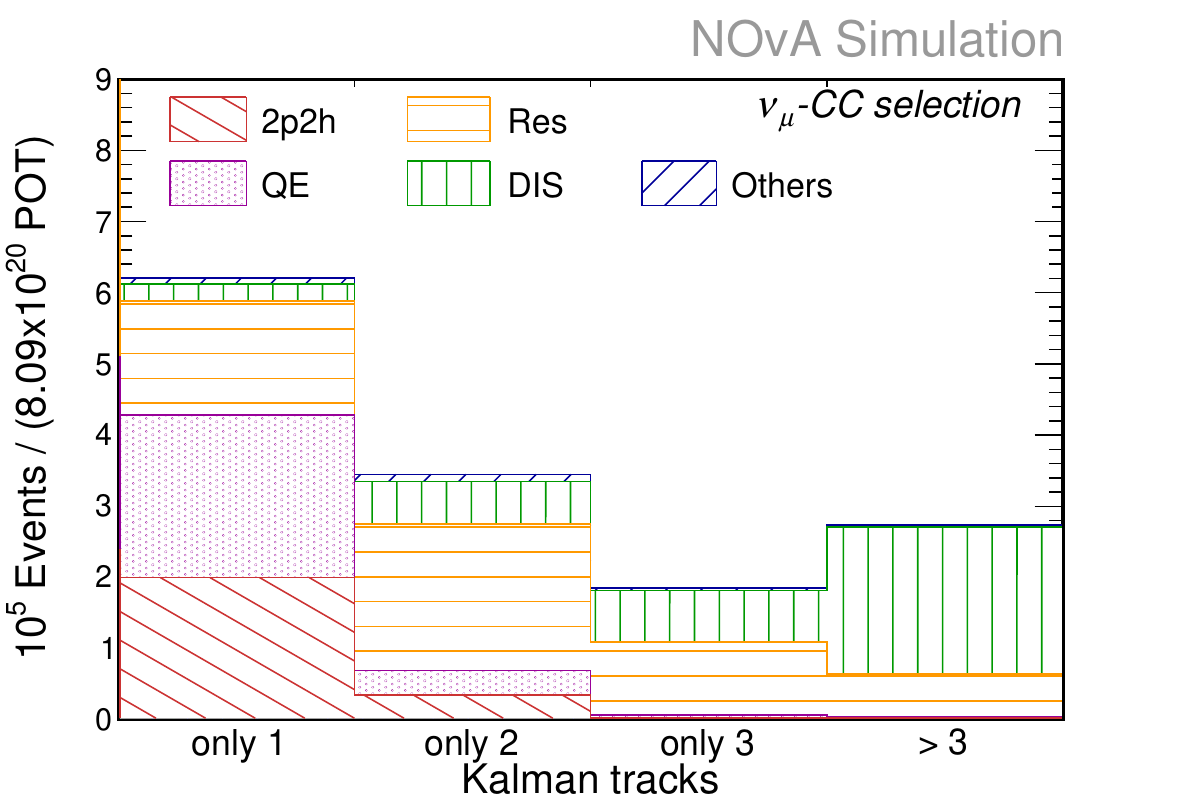} 
    \caption{Event yield per number of tracks divided by neutrino interaction types in the muon-neutrino CC sample.}
    \label{fig:ntracks}
\end{figure}

This analysis aims to select a sample of muon-neutrino CC interactions in the NOvA ND with enhanced QE and 2p2h components. 
Figure~\ref{fig:ntracks} shows that the simulated Res and DIS processes produce most of the events with more than one reconstructed track.  This is because these processes can produce final-state hadrons with enough kinetic energy for the reconstruction algorithm to produce one or more tracks. Therefore we select events with only a single track (the muon candidate). This selection represents 46.8\% of the inclusive muon-neutrino sample.

\begin{figure}[!thp]
    \centering
    \includegraphics[width=0.45\textwidth]{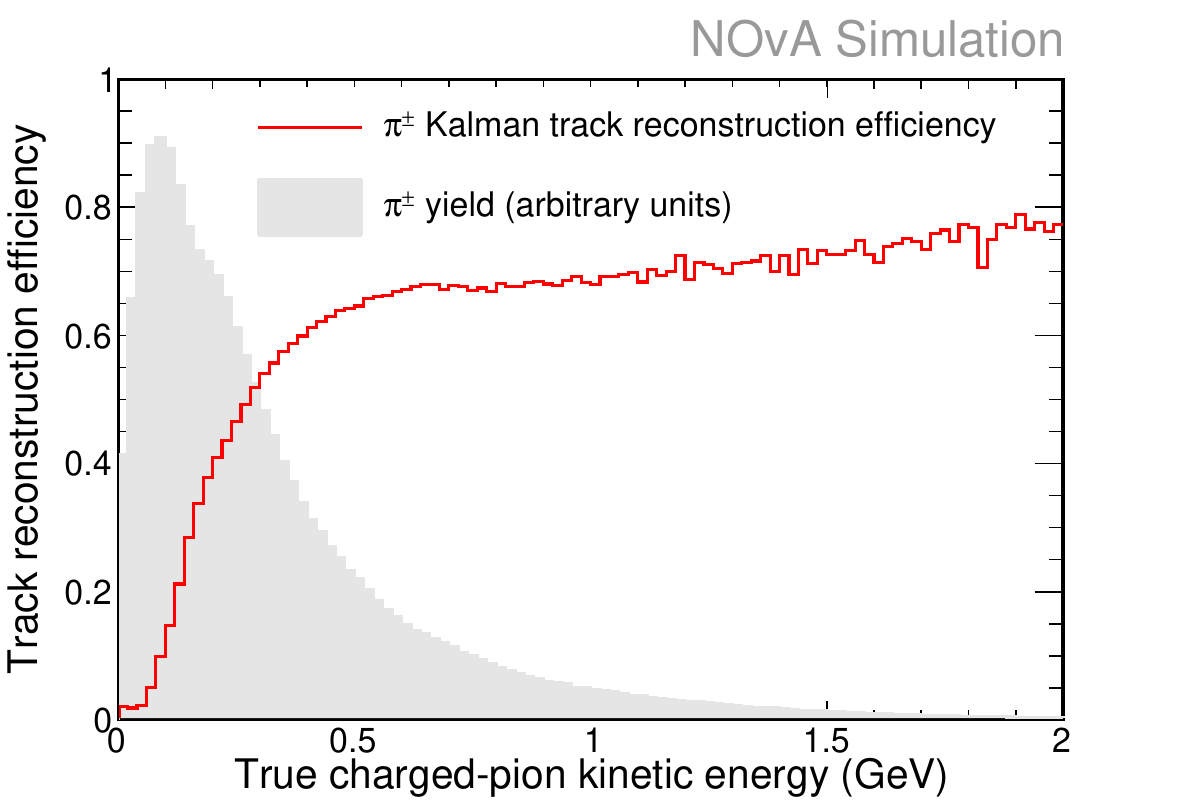}
    \caption{Track reconstruction efficiency for charged pions in the muon-neutrino selected sample overlaid with the pion energy distribution.}
    \label{fig:pireco}
\end{figure}

The single-track event selection includes interactions with hadrons in the final state that are below the track reconstruction energy threshold.  For example, Fig.~\ref{fig:pireco} shows the true charged pion yield (in solid gray) and the pion track reconstruction efficiency (solid red histogram) for simulated events with an identified and contained muon. The pion reconstruction efficiency increases with kinetic energy and reaches a plateau around \SI{0.5}{GeV}. The sharp rise in efficiency is a result of the minimum number of hits required by the track reconstruction algorithm. Simulation studies have shown that the plateau is a result of charged pion interactions in the detector medium, where the pion produces a hadron shower and the track reconstruction algorithm fails.  Since the event selection does not identify hadrons in the final state by any other means beyond the number of reconstructed tracks, we define the signal with final-state hadron kinematic limits.  A study was conducted by scanning across possible proton and pion energies and calculating the fractional uncertainty of the total cross section, with the proton and pion energy limits in the signal definition modified accordingly at every step. The  minimal fractional uncertainty was found for a signal definition with a maximum proton kinetic energy of \SI{250}{MeV} and maximum pion kinetic energy of \SI{175}{MeV}.  Adopting these limits as the signal definition, the signal selection efficiency, with respect to the total simulated signal events with vertices in the fiducial volume, is 21\%, and the purity is 80\%.  The dominant source of the selection inefficiency is the event containment requirement, and the dominant source of the selection impurity comes from events with hadrons above these limits where the reconstruction failed to find a second track.  Our simulation predicts 620,000 signal events in our final selection.  

\section{Cross-Section Extraction Method}
\label{sec:xsextration}

The main result of this analysis is the flux-integrated double-differential cross section with respect to the muon kinetic energy (\tmu) and the cosine of the muon angle with the NuMI beam direction (\cosmu). As in Ref.~\cite{bib:NOvANuMuCCInc}, the entire analysis is performed in three dimensions, which are the muon kinematic variables and the available energy (\eav), and is then integrated over \eav:

\begin{widetext}
\begin{equation}
\left(\frac{d^2 \sigma_\mathrm{incl}}{d\cos\theta_\mu \, dT_\mu}\right)_i = \frac{1}{N_\mathrm{target}\,\phi\;}\sum_{\eav}\left(\frac{\sum_j U^{-1}_{ij}[N_\mathrm{sel}(\cos\theta_\mu,T_\mu,\eav)_j \, P(\cos\theta_\mu,T_\mu,E_\mathrm{eav})_j]}{\epsilon(\cos\theta_\mu,T_\mu,\eav)_{i}\,\Delta \cos\theta_{\mu_i}\, \Delta T_{\mu_i}} \right)   \,.
\label{eq:xsec_doublediff}
\end{equation}
\end{widetext}
The available energy is the sum of electron, proton, charged pion, and kaon kinetic energy, plus neutral pion and photon total energy. For hyperons, the total energy minus the nucleon mass is added; for antinucleons it is the total energy including rest mass.  Neutron energies are excluded since the efficiency of associating energy deposits from neutron scattering with the reconstructed event is very low.  \eav is designed to correspond to the hadronic energy that can be reliably observed in the detector with minimal model dependence. In this analysis, \eav serves as a proxy of the energy of the final-state hadronic system.  The reconstruction of \eav maps the observed visible energy of simulated events not associated with the reconstructed muon to the true \eav.  The procedure of applying efficiency and purity corrections as a function of the observed final-state muon kinematics and \eav  reduces potential bias from modeling of the final-state hadronic system. We use four bins of \eav, with bin boundaries $(0, 0.15, 0.30, 0.60, 120)$ GeV, where the last bin ensures events in the tail are included in the migration matrix.

We also present single-differential cross sections in the derived variables neutrino energy (\Enu) and square of the four-momentum transfer (\Qsq). These cross sections are limited to the phase space of the muon kinematic measurement, as described below. \Enu and \Qsq are determined from a combination of the reconstructed muon energy and the visible calorimetric energy, the latter also being an input to \eav.

The muon energy is estimated by track length. Simulation studies demonstrate a muon energy resolution of approximately 4\% and the muon angle resolution is below $4^{\circ}$.  The resolution in \eav is energy dependent, ranging from 50 to 80 MeV for signal events. The 115 analysis bins shown in Table~\ref{tab:binning} are chosen according to these resolutions and the expected statistical and systematic uncertainties in the measurement (see Sec.~\ref{sec:uncertainties}).

\begin{table}[!htbp]
\begin{center}
\caption{Muon kinematic binning structure for the analysis results. 115 bins in total are used in the analysis.}
\begin{tabular}{c c c }
\hline
\multirow{ 2}{*}{ \cosmu range } & Number of  & \multirow{ 2}{*}{\tmu range (GeV)}\\
 & \tmu bins &  \\
\hline 
0.50--0.68  &   2 &  0.5--0.7 \\
0.68--0.74  &   2 &  0.5--0.7 \\
0.74--0.80  &   4 &  0.5--0.9 \\
0.80--0.85  &   6 &  0.5--1.1 \\
0.85--0.88  &   7 &  0.5--1.2 \\
0.88--0.91  &   8 &  0.5--1.3 \\
0.91--0.94  &  12 &  0.5--1.7 \\
0.94--0.96  &  15 &  0.5--2.0 \\
0.96--0.98  &  19 &  0.5--2.4 \\
0.98--0.99  &  20 &  0.5--2.5 \\
0.99--1.00  &  20 &  0.5--2.5 \\
\hline
\end{tabular}
\label{tab:binning}
\end{center}
\end{table}

\begin{figure*}[htbp]
    \centering
    \includegraphics[width=.99\linewidth]{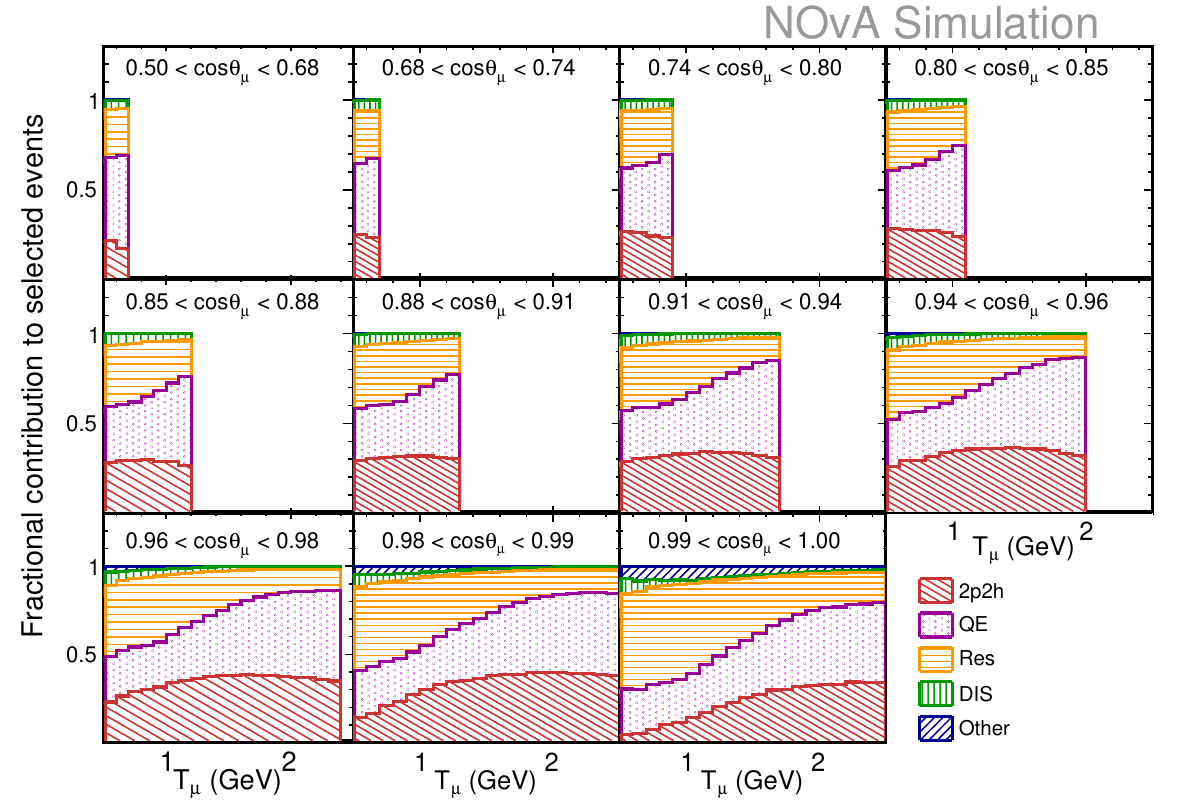}
    \caption{Fractional contribution of each interaction mode of the selected event sample in \cosmu slices using the NOvA tune v1. }
    \label{fig:RatioRecoSel}
\end{figure*}

Figure~\ref{fig:RatioRecoSel} shows \tmu distributions for selected events, one \cosmu slice per panel, broken down by the predicted fractional contributions of each interaction mode (according to the NOvA tune v1). In all slices, most interactions are from the QE, 2p2h, and Res modes. The DIS component is small and does not influence the rest of the analysis. In most of the bins, Res and 2p2h have equal contributions, except for bins with the most forward-going muons at higher kinetic energies, for which the 2p2h fraction is enhanced with respect to the Res fraction.
We identify regions in which the 2p2h component is enhanced and represents at least 35\% of the total selected events. These regions are defined as: 
     0.91~$<$~\cosmu~$<$~0.98 and \tmu~$>$~1.2 GeV; 
     0.98~$<$~\cosmu~$<$~0.99 and \tmu~$>$~1.4 GeV; 
     0.99~$<$~\cosmu~$<$~1.00 and \tmu~$>$~1.6 GeV.

The background is composed mainly of events with at least one pion or proton above the signal definition limits. Neutral-current, muon-antineutrino, electron-neutrino, and electron-antineutrino CC interactions sum to a few percent of the total prediction. The background is estimated using simulation and is subtracted from the selected sample to obtain the measured signal events. 
The purity values range from 60\% to 90\% with higher purity at forward angles, where the Res and DIS contributions are smaller.

\begin{figure*}[htbp]
    \centering
    \includegraphics[width=.45\linewidth]{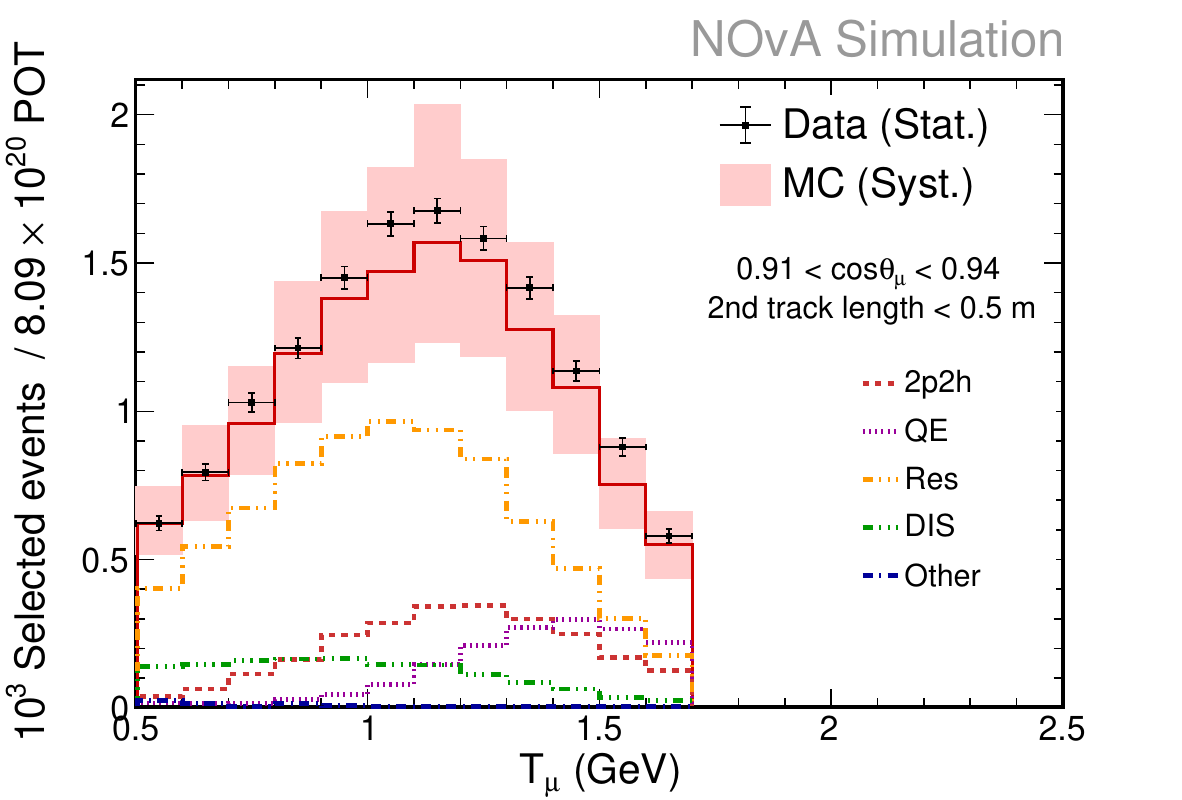}
    \includegraphics[width=.45\linewidth]{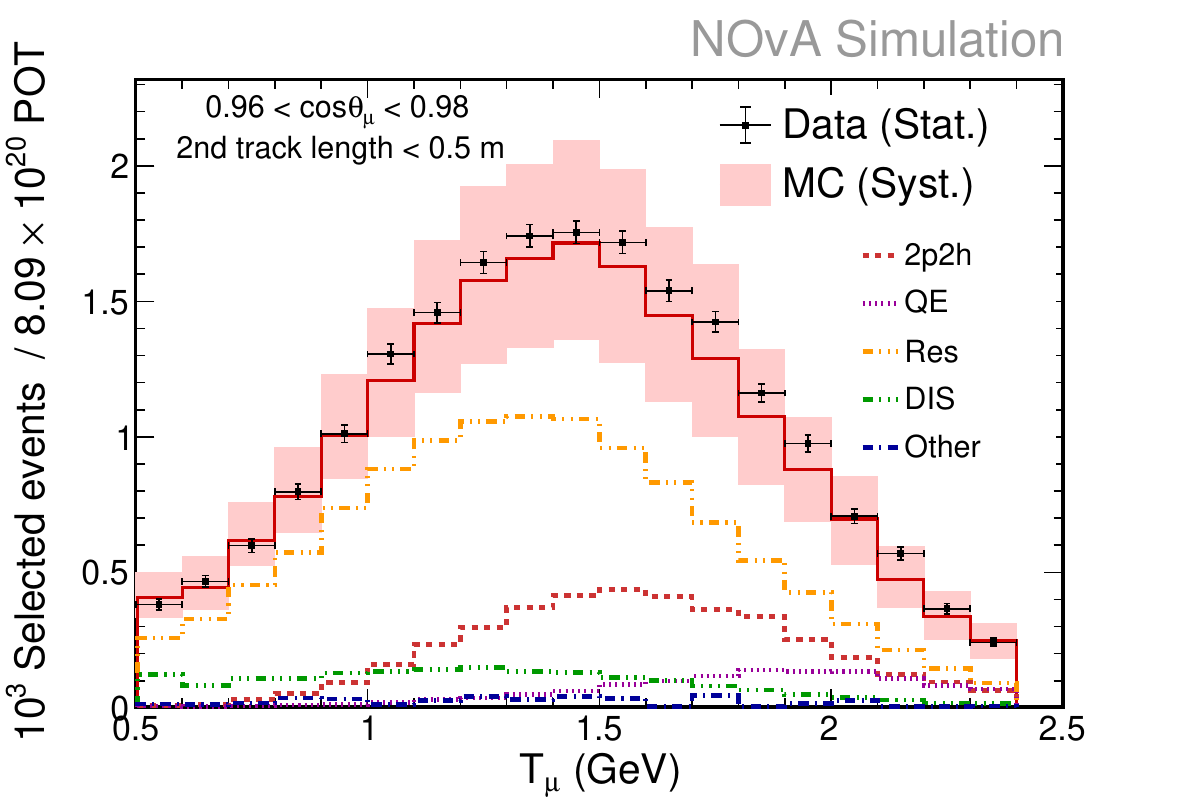}
    \caption{Two representative \cosmu slices of the sideband sample: 0.91--0.94 (left) and 0.96--0.98 (right). The sideband sample requires a muon track and one more track. It is also required that the second track have a shorter length where the tracking is highly inefficient.}  
    \label{fig:side}
\end{figure*}

The purity estimation was validated by sideband studies using events with two tracks (the muon candidate plus one additional track) where the second track has a short (less than \SI{50}{cm}) length. Such short tracks are a category for which the trackfinding algorithm is less efficient. These events were close to being included in the analysis selection. The analysis backgrounds dominate this sample. As an example, Fig.~\ref{fig:side} shows the muon kinematic distributions of two representative \cosmu slices of the sideband region.  The simulation is broken down by interaction mode and shows that resonant interactions are dominant in this sideband.  All data points fall within the systematic error band and are close to the simulation central value. 

After applying the purity corrections, we use the unfolding technique developed by D'Agostini~\cite{bib:DAgostini} and implemented by RooUnfold~\cite{bib:roounfold} to correct the smearing between bins due to detector and reconstruction effects and map the observables from the reconstructed to the true space.  The degree of smearing is small: only 0.46\% of the off-diagonal bins in the migration matrix contain counts exceeding 20\% of that of the corresponding diagonal bin. Fake data studies were performed to validate the unfolding procedure. One fake data study used five different systematically shifted $Q^2$ distributions of the signal events, where the \Qsq dependence was modified by up to 5\% at \Qsq=0 and up to 30\% at \Qsq=1.5 GeV$^2$.  Another fake data study used systematically shifted 2p2h models, where the NOvA-tune 2p2h model was reweighted to match the SuSAv2 and MINERvA models.
We use the minimal Mean Square Error~\cite{cowan1998statistical} of the unfolded distribution of fake data with respect to the true distribution
to optimize the number of iterations. In all cases, performing three iterations was found to be optimal for this analysis, and the unfolded distributions statistically agreed within the uncertainties of the underlying model.

\begin{figure*}[htbp]
    \centering
    \includegraphics[width=.75\linewidth]{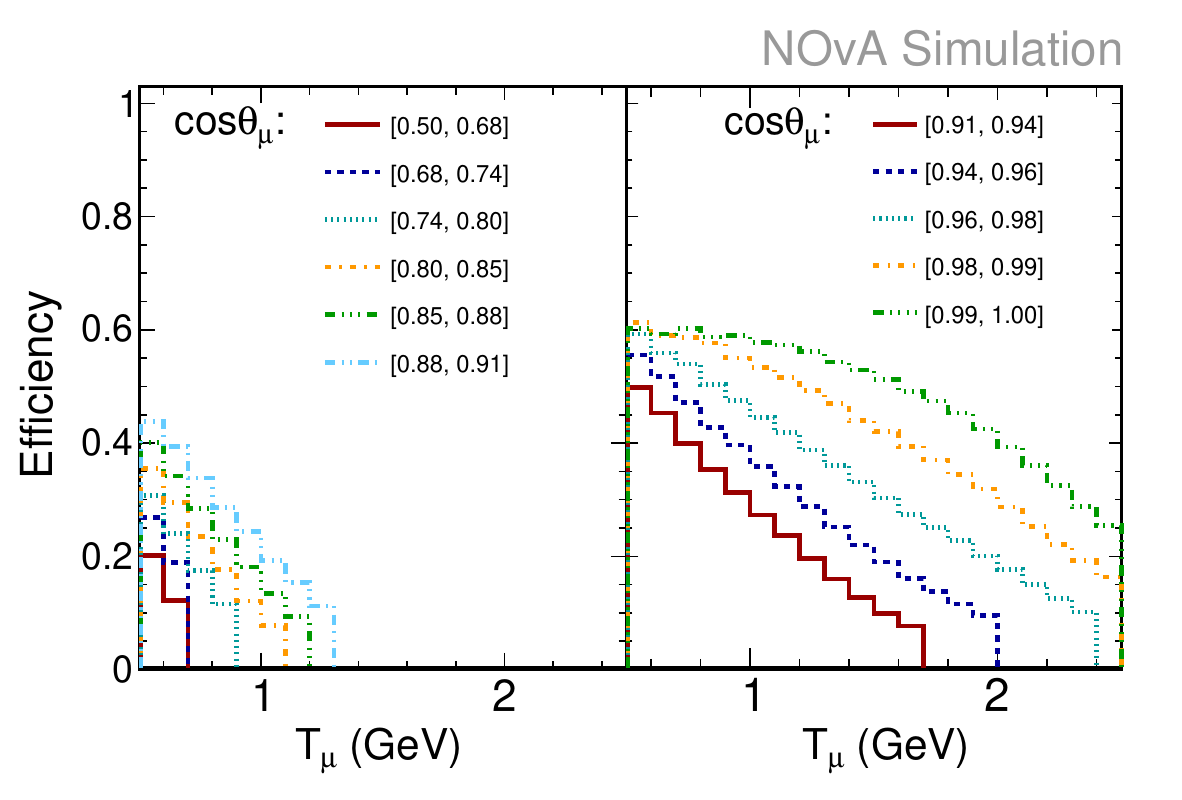}
    \caption{Efficiency distributions for each \cosmu slice. The values decrease as the muon kinetic energy increases. Conversely, the efficiency increases as more forward-going muons are selected. 
    This pattern results from the containment requirement.}
    \label{fig:Eff}
\end{figure*}

The selection efficiency vs.\ \tmu in slices of \cosmu is shown in Fig.~\ref{fig:Eff}. The shape and value of the efficiency arise primarily from the containment requirement. It sharply decreases for higher energy as the muon tracks become less likely to be contained in the detector. The efficiency decreases at larger angles because muon tracks with larger angles can escape more easily from the detector's sides.

The integrated neutrino flux is $9.64 \times 10^{16}$ per \SI{}{m^2} with a 11.4\% error. The fiducial volume consists of 64.6 tons of scintillator (62.75\%), PVC (36.54\%), and glue (0.71\%).  The number of nucleon targets in the fiducial volume is estimated as $3.89 \times 10^{31} \pm 0.33\%$.

\section{2p2h models}
\label{subsec:2p2hmodels}
Since a fully relativistic 2p2h model calculation that considers all relevant diagrams is challenging, and existing predictions disagree significantly, we expect the 2p2h component of the signal to be the most important aspect to study. We therefore analyze our data by comparing the measured cross sections to simulated ones that use differing 2p2h models.
These simulations are implemented by removing the 2p2h fit-to-NOvA-data component from the NOvA tune v1 and reweighting the untuned 2p2h component to the desired model. This procedure is performed in the phase space of the energy transfer, $q_0$, vs.\ the magnitude of the three-momentum transfer, $q_3$. Six 2p2h models were included in this comparison to sample the range of theoretical treatments of the 2p2h process:

\begin{enumerate}
\item the microscopic first-principles \valencia model~\cite{bib:Valencia1,bib:Valencia3};
\item the microscopic first-principles model from the SuSA group(\susa)~\cite{bib:SuSAv2};
\item the GiBUU 2021 2p2h model~\cite{bib:GiBUU1,bib:GiBUU2};
\item the Empirical model implemented in GENIE~\cite{bib:MECModels};
\item the experimentally based MINERvA tune of the \valencia model (MnvTune-v1.2)\cite{bib:MINERvALELowq3}; and
\item the NOvA tune v1 to GENIE v2.12.2~\cite{bib:NOvATune2020}.
\end{enumerate} 

The first three models are based purely on theory and phenomenology, and the last three are experimentally-based empirical predictions.  The theory and phenomenology-based models incorporate 2p2h interactions within a unified framework for electroweak interactions. The \valencia model is based on a many-body expansion of the gauge boson absorption modes and includes nuclear effects such as RPA and short-range nucleon-nucleon correlations. When calculating the 2p2h cross sections, the model incorporates the calculation of the longitudinal and transverse nuclear response functions. A kinematic cutoff at 1.2 GeV momentum transfer is implemented to enable the model's application at energies up to 10 GeV.

The SuSAv2 model, a comprehensive extension of the original SuperScaling Approach (SuSA)\cite{bib:SuSA}, incorporates 2p2h in its fully relativistic framework (relativistic mean fields). While both SuSAv2 2p2h and the Valencia model share a common RFG-based 2p2h calculation, they differ in handling the $\Delta$-resonance propagator. SuSAv2 uses only the real part to avoid double counting effects between the 2p2h and inelastic regimes, achieving good agreement with electron scattering data. Conversely, the Valencia model includes both real and imaginary parts, accounting for higher-energy resonance exchange. 

GiBUU incorporates 2p2h interactions in its quantum-kinetic transport theory of lepton--nucleus reactions. This model assumes that 2p2h contributions are predominantly transverse, leveraging an empirical structure function derived from electron scattering data. It includes effects from short-range nucleon-nucleon correlations and RPA. The model is based on the theoretical framework established by Walecka and others~\cite{bib:walecka1,bib:walecka2}, which connects electron response to neutrino response and assumes that the longitudinal nuclear response is negligible.

The Empirical model, first implemented in GENIE, characterizes the 2p2h differential cross section as a Gaussian distribution as a function of \Qsq between the QE and Delta resonance peaks. This approach addresses the observed excess in electron scattering data between these two peaks. The model emphasizes the transverse nuclear response over the longitudinal response using the Sachs magnetic form factor to define neutrino and electron interaction cross sections. The 2p2h strength is extracted from electron scattering data and applied to predict neutrino cross sections.

The MINERvA experiment uses a 2p2h tune of the \valencia 2p2h model implemented in GENIE as the default simulation to match its observed data better. The tune is performed in the energy transfer and three-momentum transfer phase space to adjust the additional strength required, especially at moderate three-momentum transfer, particularly in the region between the QE and Delta resonance peaks. This model enhancement addresses the \valencia model's underestimation of event rates in this region and significantly improves the agreement with MINERvA's neutrino and antineutrino data.

\begin{figure*}[htbp]
    \centering
    \includegraphics[width=.85\linewidth]{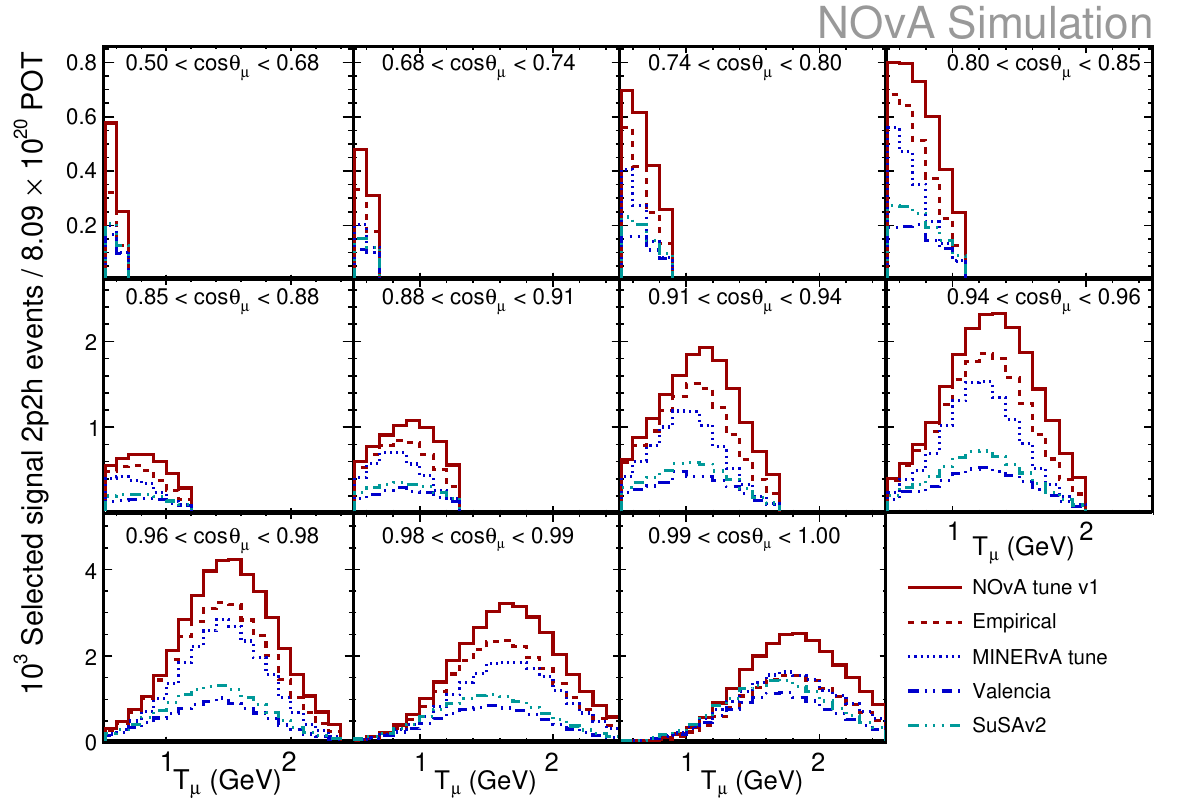}
    \includegraphics[width=.85\linewidth]{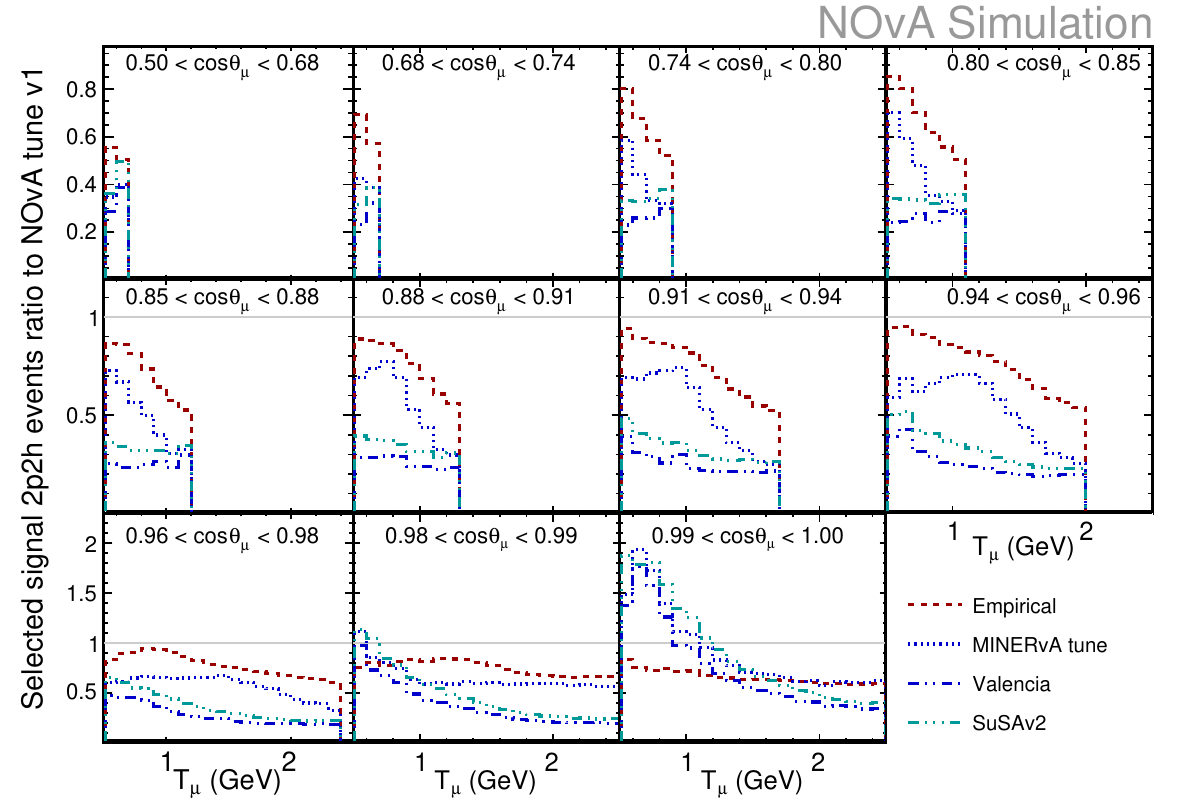}
    \caption{2p2h component in the selected signal distribution using various 2p2h models (NOvA tune v1, Empirical, MINERvA tune (MnvTune-v1.2), \valencia, and \susa) is shown in the top panels. Ratios to the NOvA tune v1 are shown in the bottom panels.
    }
    \label{fig:RatioMECs}
\end{figure*}

Figure~\ref{fig:RatioMECs} shows the 2p2h components of the signal in this analysis for the \valencia, \susa and Empirical models, as well as those of the MINERvA and NOvA tunes to GENIE. GiBUU is not included as our base simulation is based on GENIE.
There are significant variations in the predicted levels and shapes of 2p2h contributions to the final-state muon kinematics, including factors of 2--3 in the predicted signal event rate, and up to 200 MeV in the position of the peak muon kinetic energy.

\afterpage{\clearpage}

\section{Uncertainties}
\label{sec:uncertainties}

\begin{figure*}[htbp]
    \centering
     \includegraphics[width=.42\linewidth]{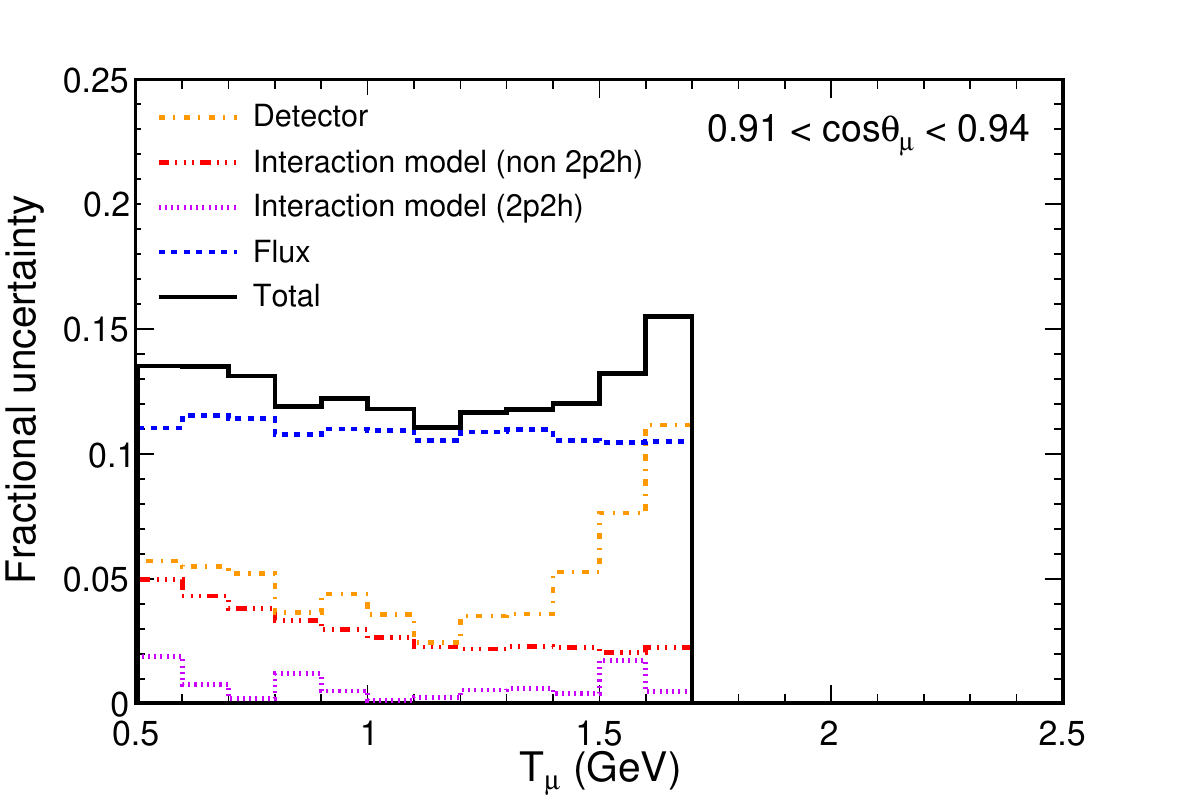}
     \includegraphics[width=.42\linewidth]{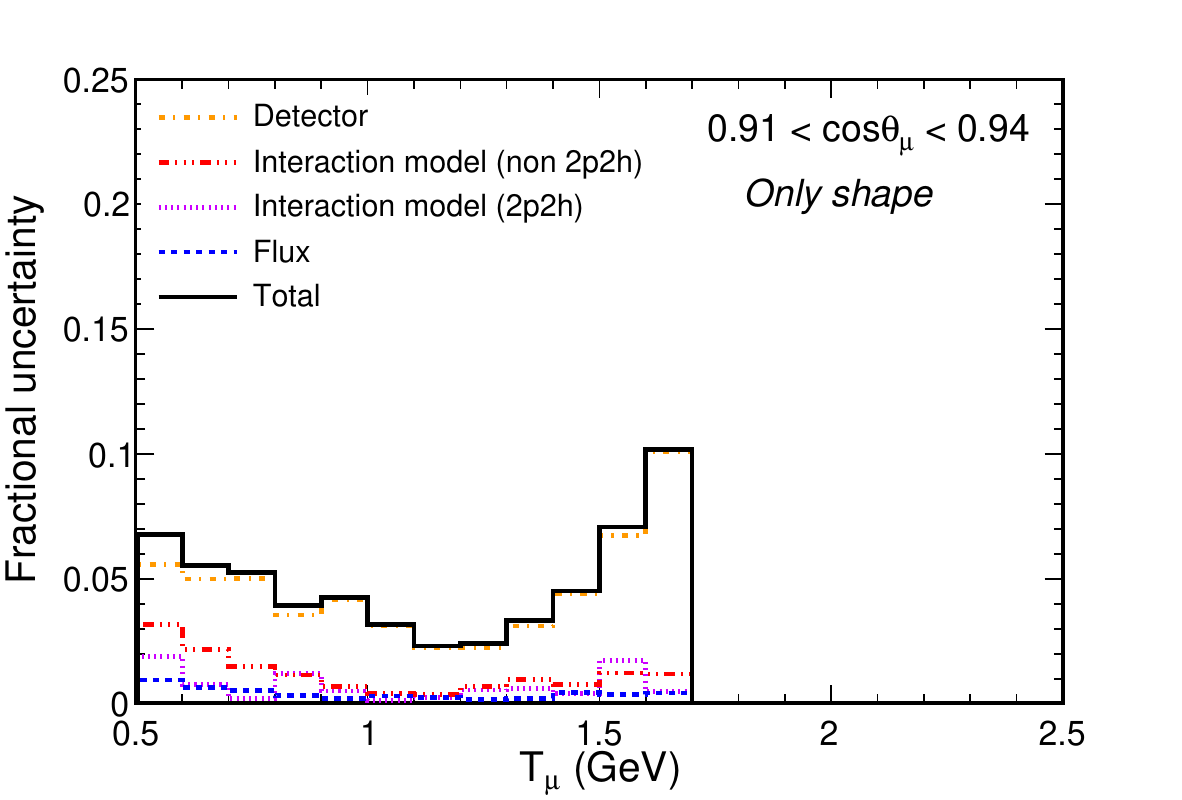}
     \includegraphics[width=.42\linewidth]{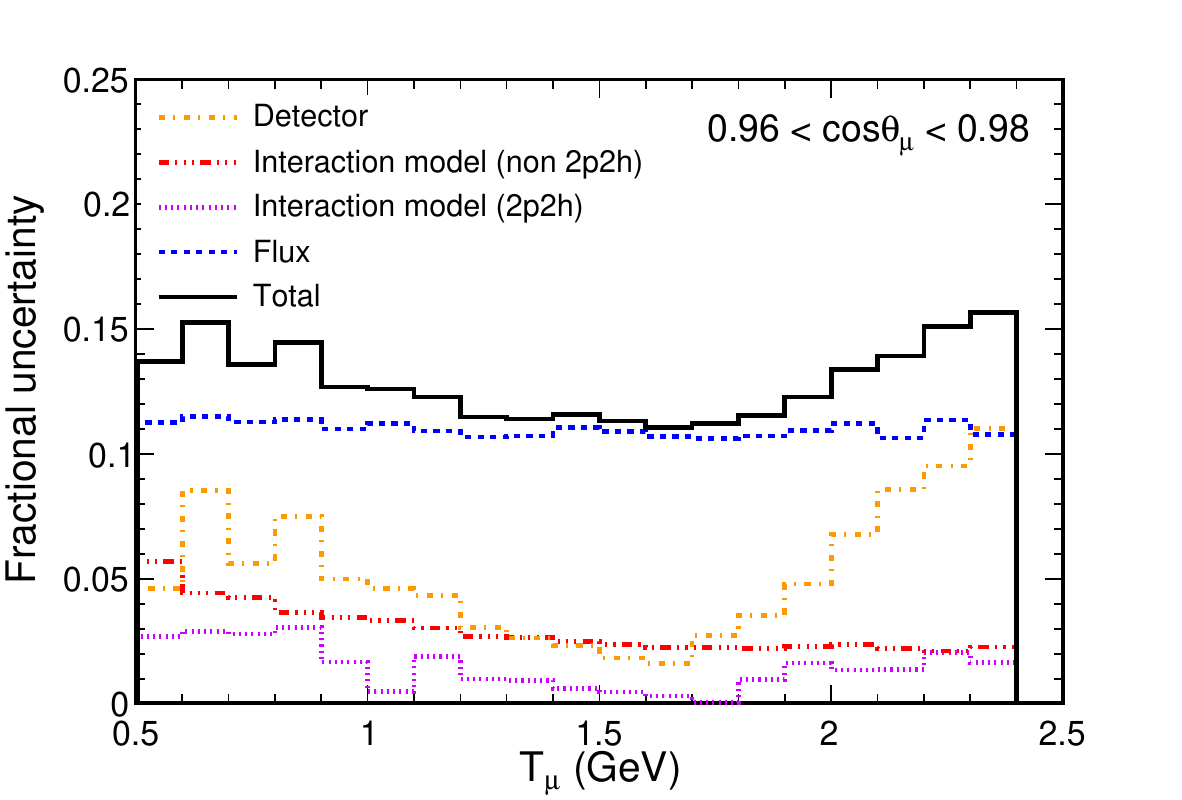}
     \includegraphics[width=.42\linewidth]{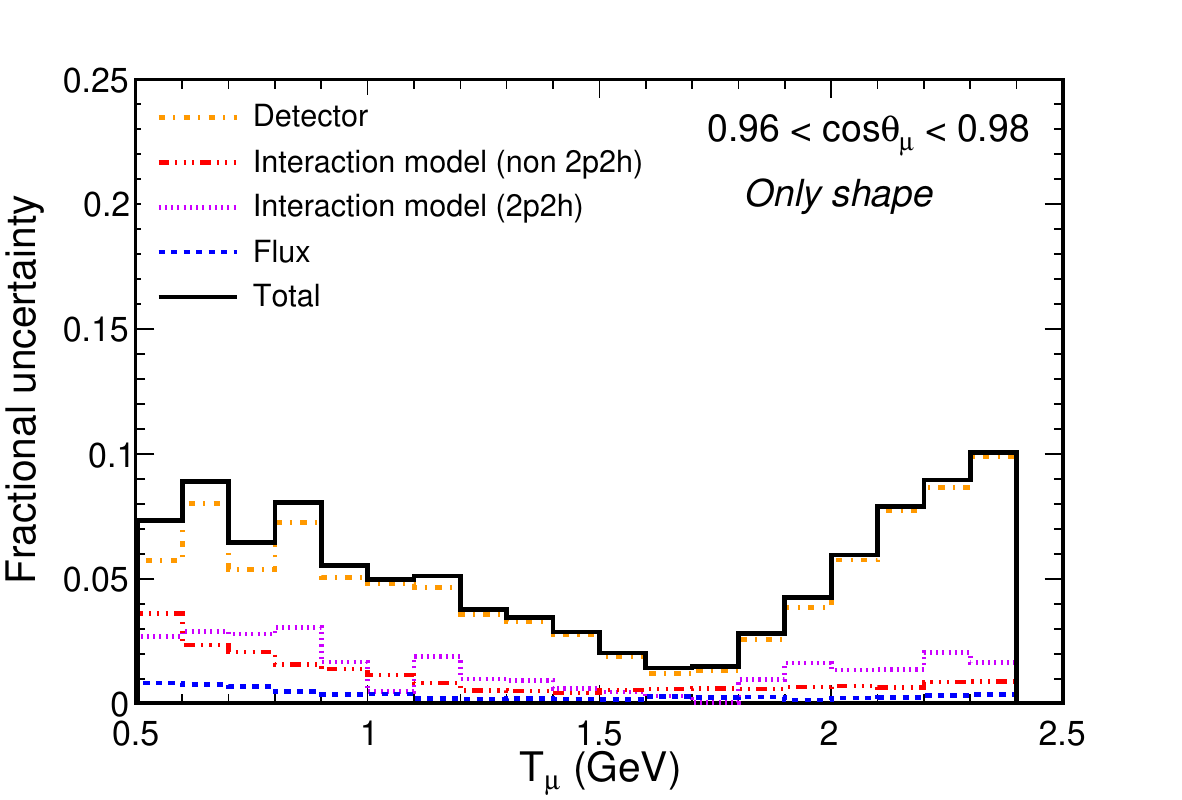}
    \caption{Fractional systematic uncertainties for the muon kinematic analysis. The left side shows the relative total (normalization + shape)value, and the right side shows the impact of removing the normalization part of the systematic source. Two representative \cosmu slices are displayed; the rest follow a similar pattern.}
    \label{fig:systmukin}
\end{figure*}

\begin{figure*}[htbp]
    \centering
    \includegraphics[width=.42\linewidth]{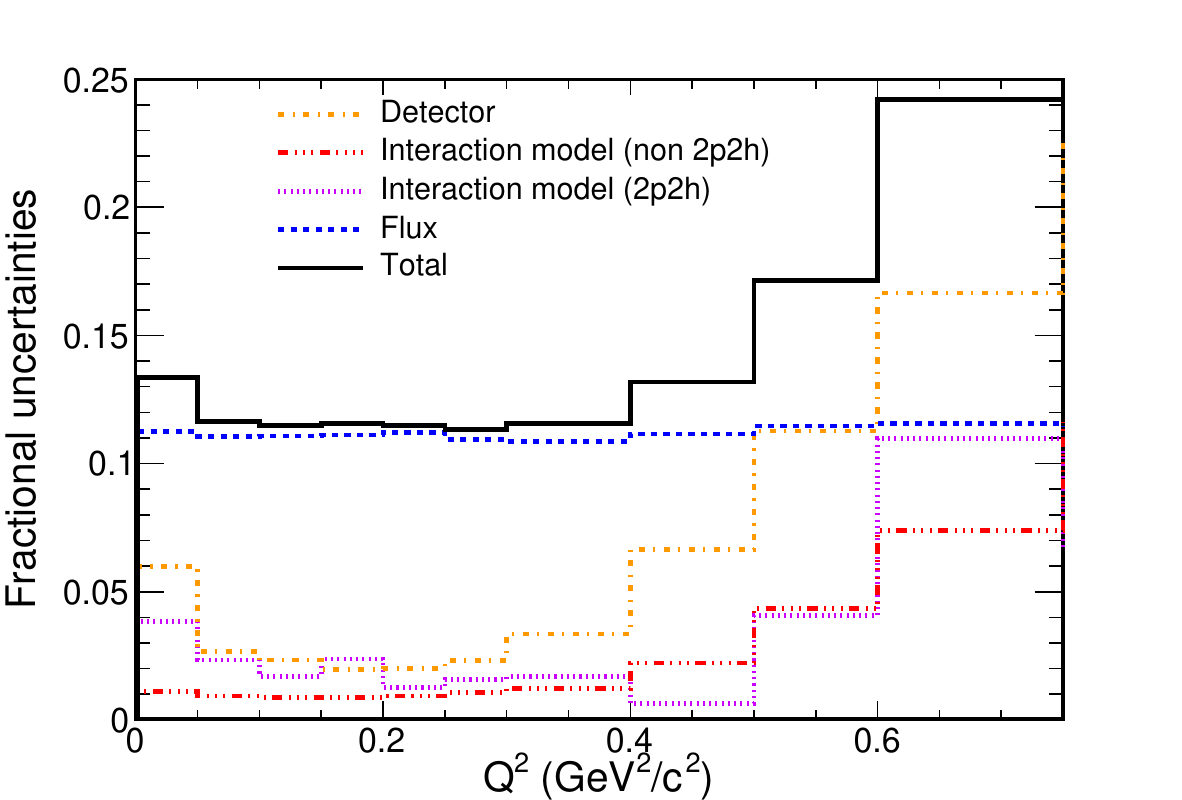}
    \includegraphics[width=.42\linewidth]{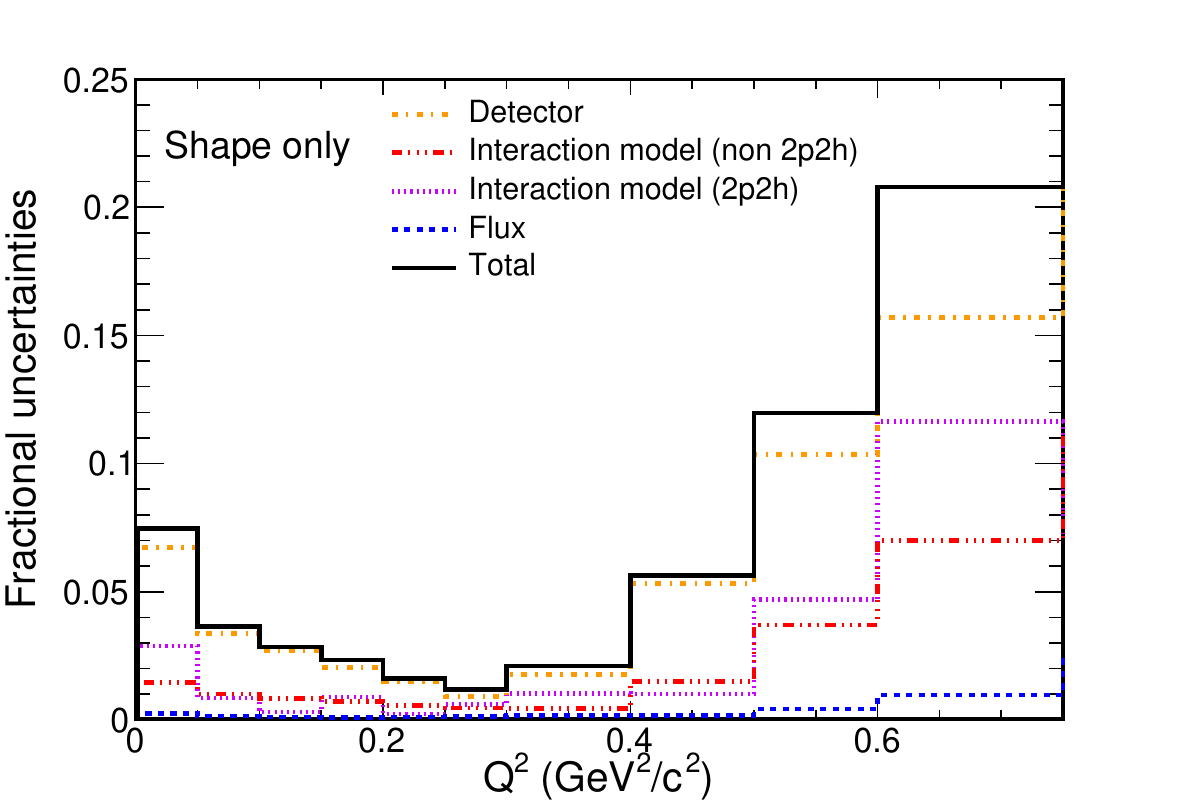}
    \includegraphics[width=.42\linewidth]{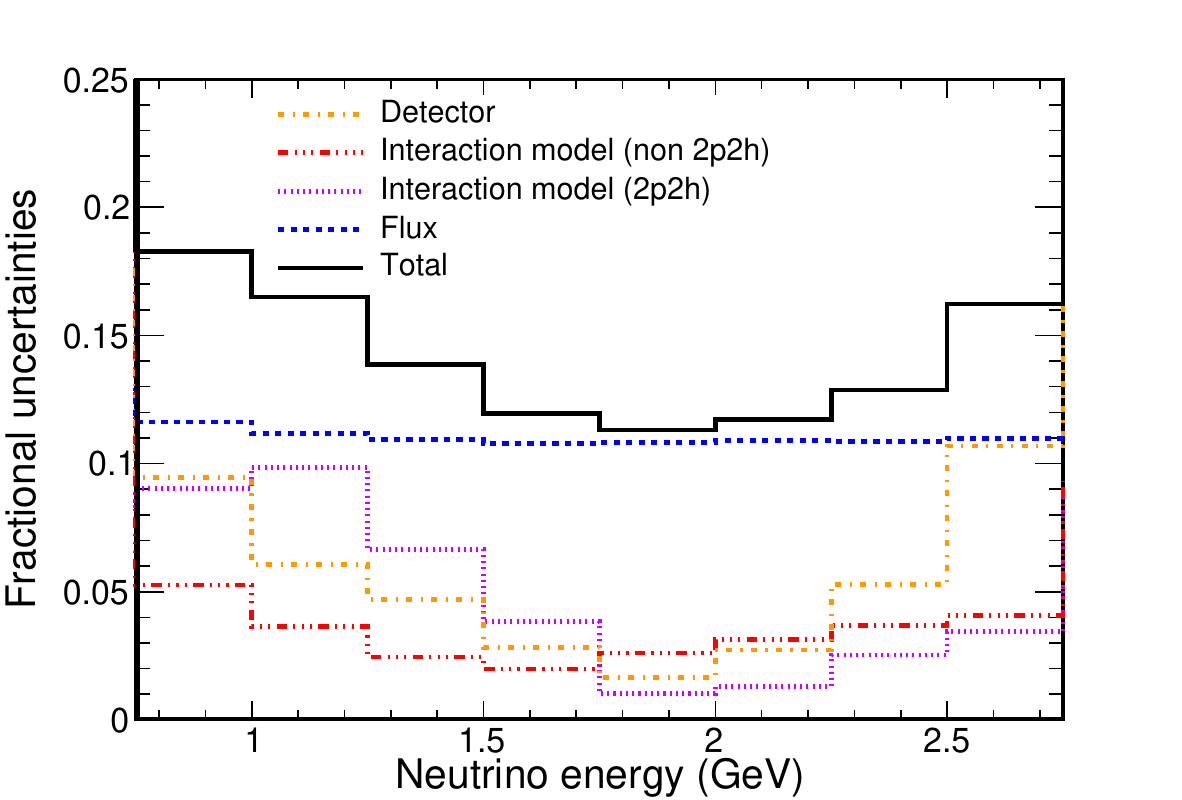}
    \includegraphics[width=.42\linewidth]{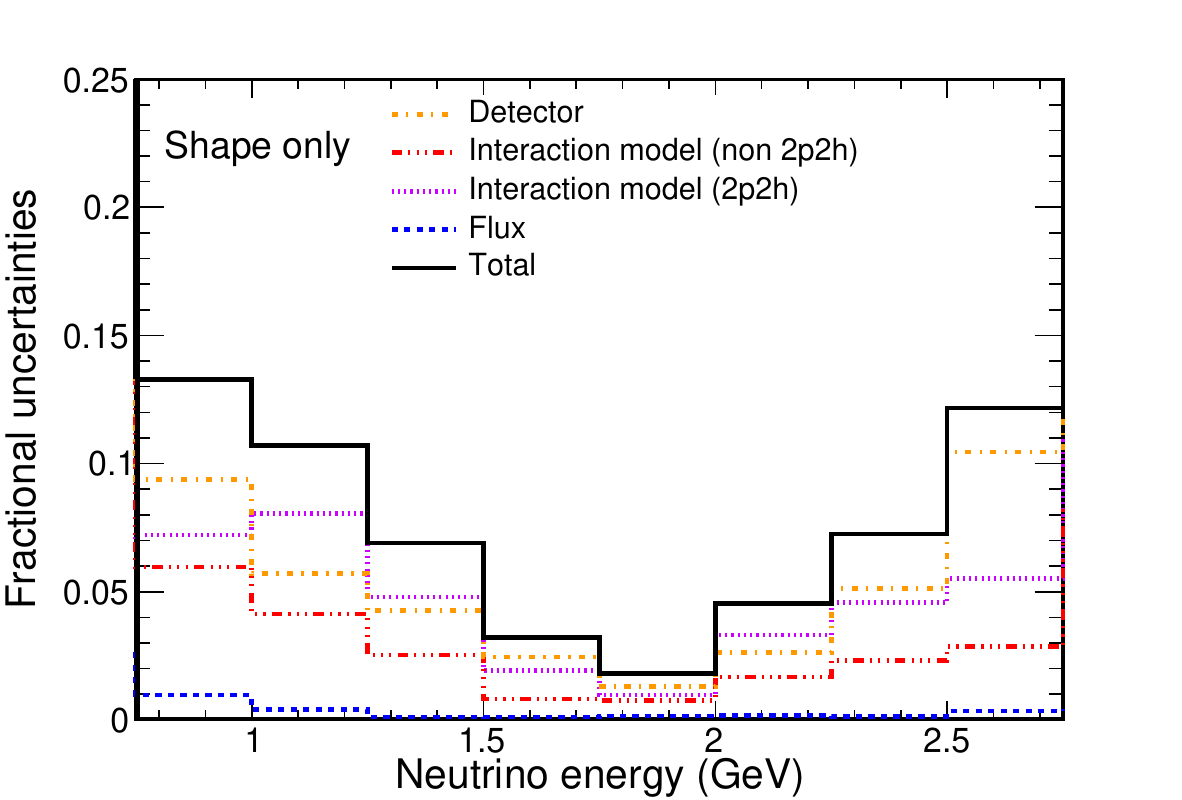}
    \caption{As in Fig.~\ref{fig:systmukin}, but for $Q^{2}$ (top) and the neutrino energy (bottom). The left side shows the absolute value, and the right side shows the impact of removing the normalization part of the systematic source.}
    \label{fig:syst1Dana}
\end{figure*}

A breakdown of fractional systematic uncertainties in the cross-section measurements is shown in Figs.~\ref{fig:systmukin} and \ref{fig:syst1Dana}. Figure~\ref{fig:systmukin} includes two representative \cosmu slices, 0.91--0.94 and 0.96--0.98 (other slices show similar behavior). The main sources 
arise from the flux and neutrino interaction modeling, detector response modeling, muon energy loss modeling, PVC cell misalignment, and particle transport modeling. The right side of the plots displays the uncertainties of the shape-only measurements, which are minimal around the maximum of the cross-section distribution in the slice. The foundation of our procedure is to repeat the cross-section evaluation using a large number of systematically shifted simulations to calculate a covariance matrix that fully accounts for correlations and their impact on the results. 

The dominant systematic uncertainty is the hadron production part of the flux prediction, which originates from uncertainties in the hadronic interaction cross sections of the particles produced in the NuMI beamline, as described in \cite{bib:MINERvAPPFX}. The impact is about 10\% and  almost flat in most of the distributions. This uncertainty is thus negligible in the shape-only analyses. The flux focusing uncertainties come from uncertainties in the NuMI component geometry, such as the alignment of the magnetic horns. The impact is 4\% and also predominantly a normalization effect.

Neutrino interaction uncertainties are split into two categories: those associated with the 2p2h process and those due to any other source. For the former, we recalculate the cross section replacing the 2p2h model in the NOvA tune v1~\cite{bib:NOvATune2020} with two other models: the \susa model~\cite{bib:SuSAv2} and the MnvTune-v1.2 tune~\cite{bib:MINERvALELowq3} of the \valencia model~\cite{bib:Valencia1,bib:Valencia3}. The maximum spread in the results is taken as the uncertainty for this category. For the non-2p2h part, we use GENIE tunable physics parameters and a set of other shifts from physics considerations and external data as described in Ref.~\cite{bib:NOvANuMuCCInc}. The impact on the final results is at the few-percent level. 

The detector response uncertainty comes from the calibration and light model. The calibration component accounts for the potential mismodeling of the light attenuation as the light travels through the WLS fibers from the location of the energy deposition to the readout end of the fiber and the conversion factor of the detector response to energy. To evaluate the effect of such mismodeling, a $\pm5\%$ shift is applied to this factor for each hit. The light model also accounts for Cherenkov photon emission and absorption efficiency. Energy depositions from protons are shifted down by 2.6\%, based on comparisons between observed energy depositions by muons and protons in data. The impact on the final results is at the few-percent level.

The muon energy-scale~\cite{bib:NOvAmuonEnergyScale} systematic uncertainty accounts 
for possible mismodeling of the relationship between the muon kinetic energy and the reconstructed muon range in the detector.
Shifts of the reconstructed energy are applied independently to both the portion of the track in the fully active region ($\pm$0.79\%) and the portion of the track in the muon catcher ($\pm$1.2\%). 
Small detector misalignments in the muon angle are constrained to be less than \SI{2.5}{mrad}. Together, these uncertainties produce migration across the analysis bins, modifying the values between 6\% and 10\%.

Charged-pion track reconstruction can fail when the particle undergoes inelastic scattering in the detector, therefore the rate of these failures can impact the selection efficiency and reconstructed \eav.  We apply a $\pm$20\% change in the rate of charged-pion inelastic scattering in the detector, based on the spread of hadron scattering predictions.  The resulting few-percent systematic uncertainty is small but not negligible. Other systematic sources such as the neutron transport model, the target nucleon count, the number of protons-on-target in the NuMI beamline, and reconstruction failures due to overlapping events in the detector due to beam intensity are also small, but are included in the final covariance matrix. 

The effect of statistical uncertainties is estimated by an ensemble of shifted event-count distributions (before unfolding) using  Poisson statistics. This ensemble is propagated through the unfolding procedure to recalculate the cross sections. In this way, we account for the correlations induced by the unfolding technique.  

\section{Results and comparisons}
\label{sec:results}

\begin{figure*}[htbp]
    \centering
    \includegraphics[width=.90\linewidth]{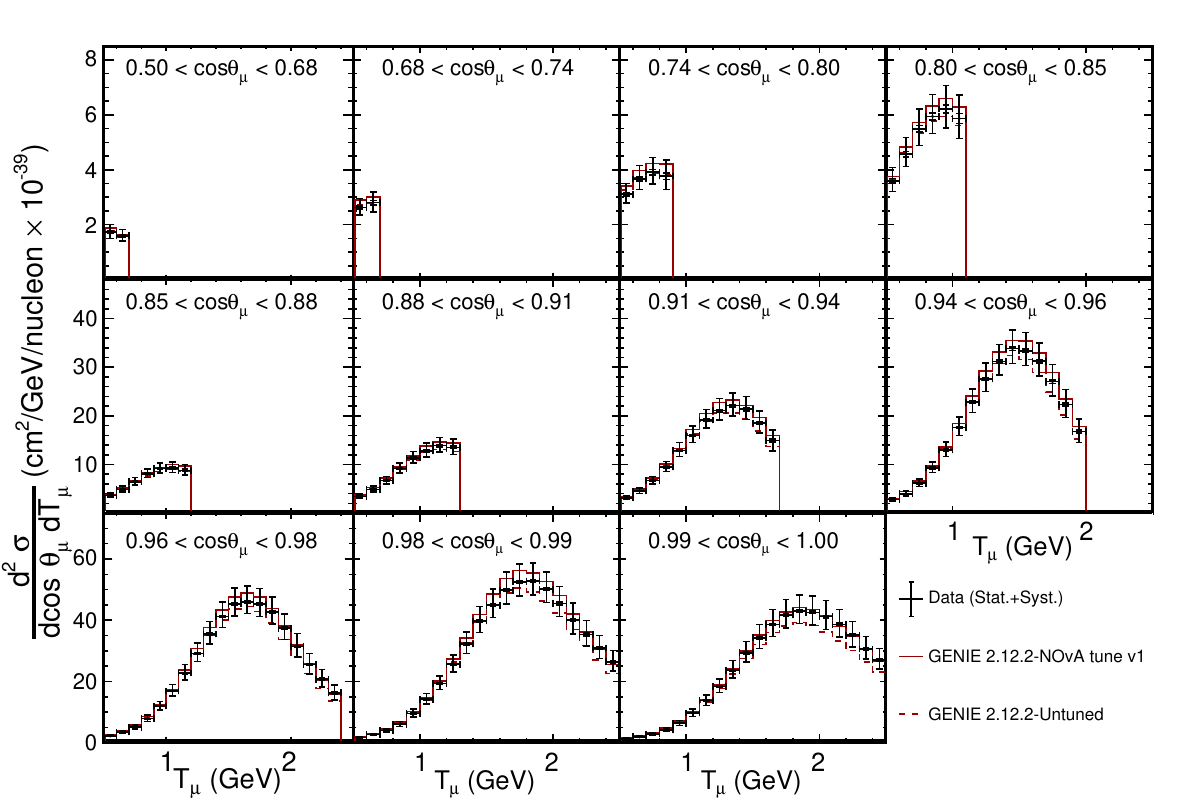}
    \caption{Extracted double-differential cross section in slices of \cosmu. The error bars represent the totaluncertainties (the inner bar represents the statistical uncertainty). The data are compared to the NOvA tune v1 (solid red) and GENIE v2.12.2 (dashed red) predictions.}
    \label{fig:XSec2D}
\end{figure*}

\begin{figure*}[htbp]
    \centering
    \includegraphics[width=.45\linewidth]{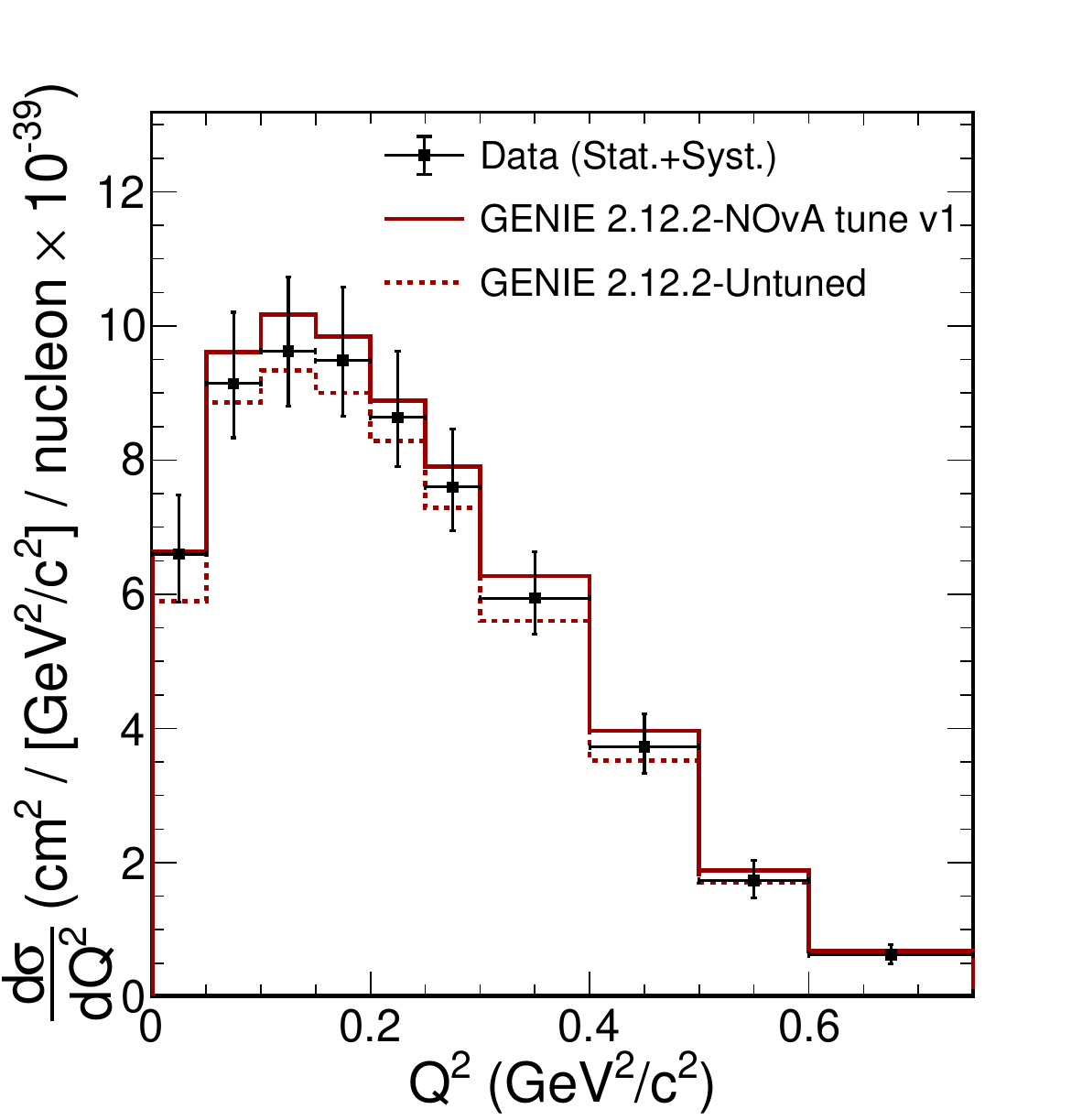}
    \includegraphics[width=.45\linewidth]{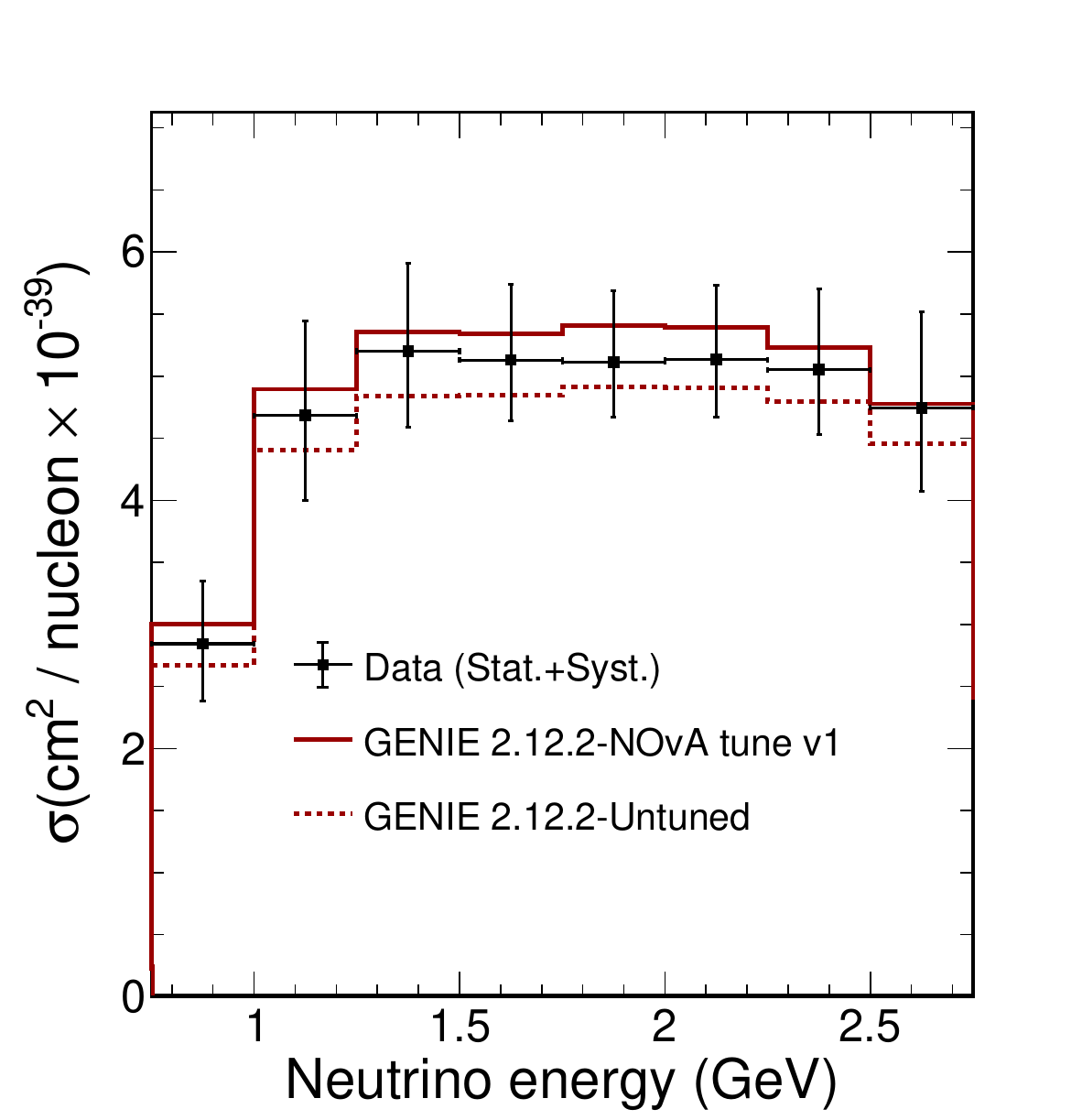}
    \caption{As in Fig.~\ref{fig:XSec2D}, but for \Qsq (left) and neutrino energy (\Enu, right).}
    \label{fig:XSec1D}
\end{figure*}

595211 data events were selected.  The double-differential cross-section measurements are presented in Fig.~\ref{fig:XSec2D} in the eleven \cosmu slices. The error bars on the data point contain statistical and systematic uncertainties. Due to the high statistics of the sample, the inner bar, representing statistical uncertainties, is almost invisible in most of the bins. We also include the predictions of the NOvA tune v1 and the untuned GENIE v2.12.2 used for the initial simulation (with Empirical 2p2h). Figure~\ref{fig:XSec1D} shows the measurement of the single-differential cross section with respect to momentum transfer (\Qsq, left) and neutrino energy (\Enu, right). 
 
\begin{figure*}[htbp]
    \centering
    \includegraphics[width=.85\linewidth]{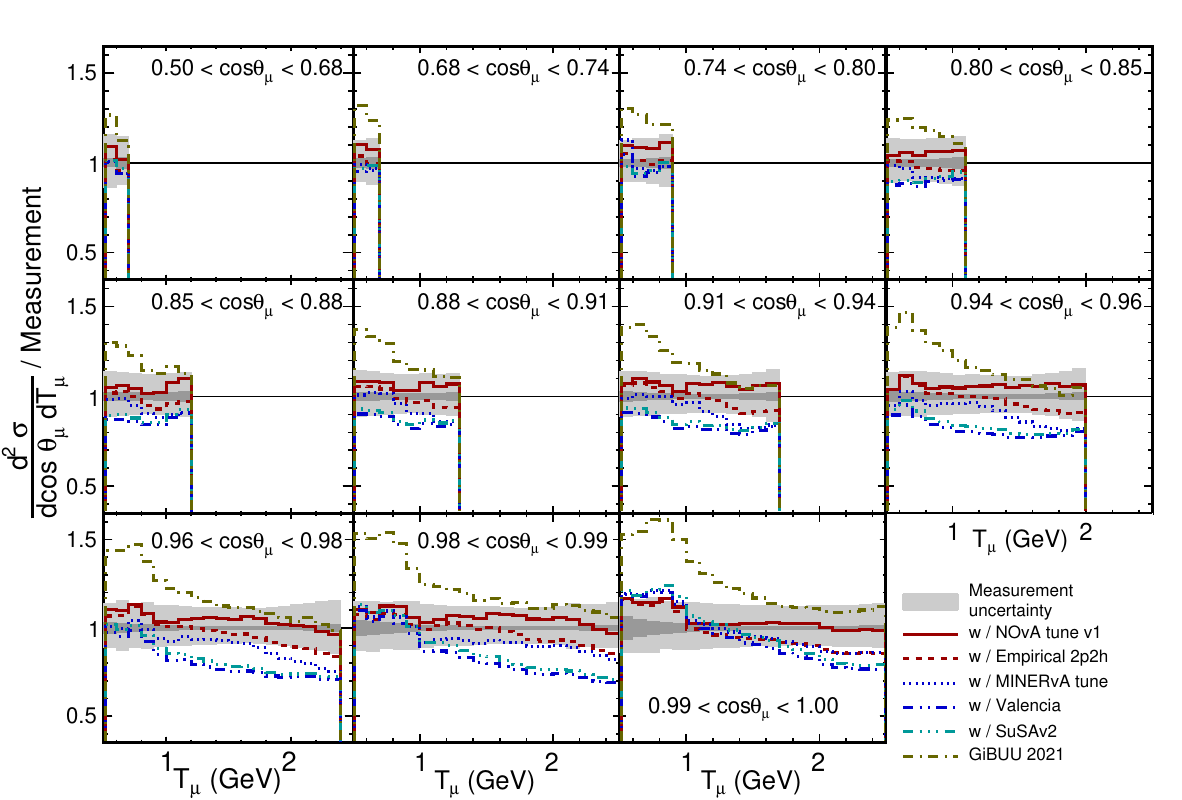}
    \includegraphics[width=.85\linewidth]{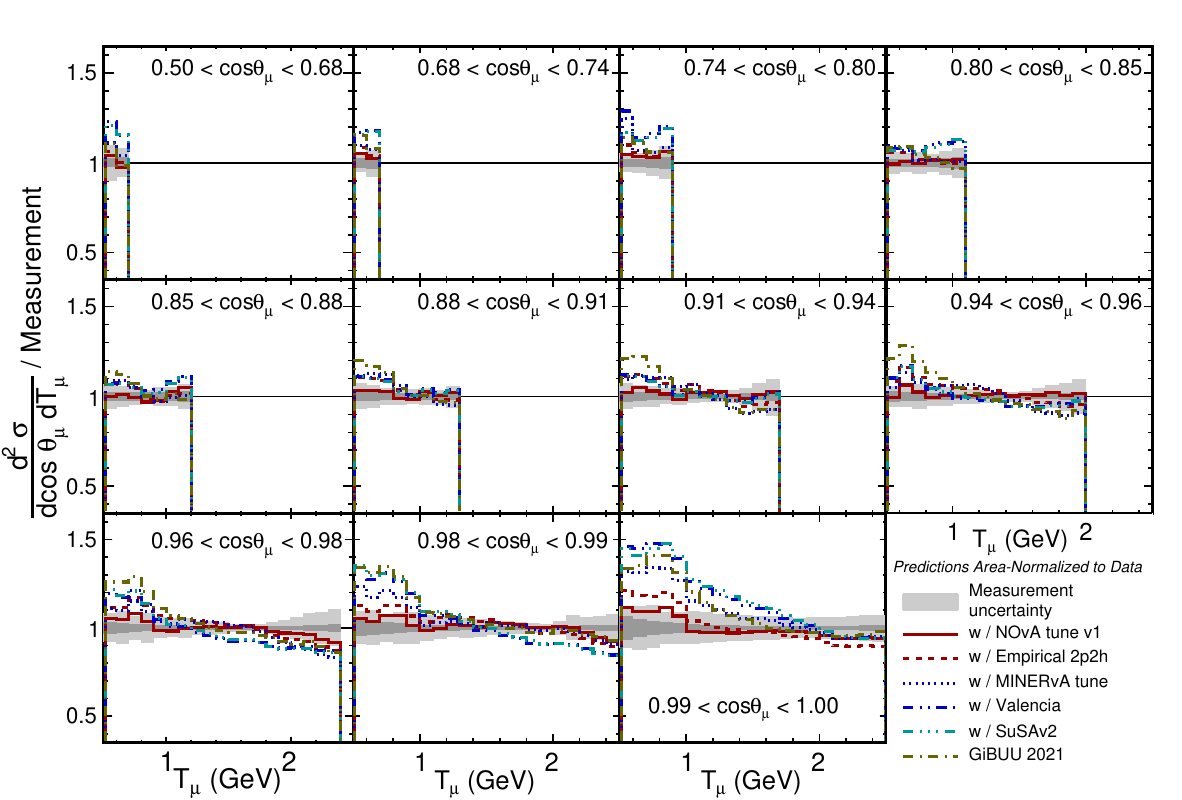}
    \caption{Ratios of predictions using different 2p2h models (Empirical, \valencia, \susa, MINERvA tune (MnvTune-v1.2) and GiBUU 2021) to our double differential muon kinematics cross-section measurement, shown in slices of \cosmu. The light (dark) gray area corresponds to the total (statistical) systematic uncertainty band. Top: ratio shown with total uncertainty; bottom: ratio shown with shape-only uncertainty and models area-normalized to data.}
    \label{fig:XSecRat2D}
\end{figure*}

Figure~\ref{fig:XSecRat2D} shows the ratio of various 2p2h simulations (Empirical, \valencia, \susa, MINERvA tune, and GiBUU 2021) to our measurement. The light (dark) gray shaded area indicates the total (statistical) uncertainty band.
The bottom panel shows shape-only comparisons.  Shape-only distributions are extracted by area-normalizing the simulation for each uncertainty to the measured cross section, thereby removing the normalization part of the uncertainty (dominated by the flux uncertainty).

None of the models reproduce the shape of the results reported in this paper. To gauge the amount of 2p2h predicted in each bin, we can refer to Fig.~\ref{fig:RatioMECs}, which includes the ratio of the 2p2h components in the selected signal distribution using models (Empirical, \valencia, \susa, and MINERvA tune) to the NOvA tune v1.

Figure~\ref{fig:XSecRat2D} shows that the NOvA tune v1 slightly overestimates the cross section in many of the bins of lower muon energy and nearly all bins of higher angles.  As shown in Fig.~\ref{fig:RatioMECs}, the MINERvA tune, Empirical, \valencia, and \susa tend to predict lower values than the NOvA tune v1, especially for higher muon energies. As a result, these predictions agree better with data for larger angles but highly underestimates the data for forward-going muons. GiBUU overestimates the data for higher angles but falls within the uncertainty for at high muon energy.
The shape-only comparisons at the bottom of Figure~\ref{fig:XSecRat2D} show that the data exhibits a steeper dependence of the cross section as a function of the energy, especially for forward-going muons.

The global $\chi^{2}$ values of the models with respect to our measured cross sections are presented in Table~\ref{tab:chi2}. This calculation uses the covariance matrix described in the previous section to account for correlated uncertainties across all analysis bins. The Empirical 2p2h has the lowest $\chi^{2}$, followed by the data-tuned models. The values in parentheses are the shape-only $\chi^{2}$ calculations. 
Since the derived variables used for the NOvA tune v1 are highly correlated with the direct observables of this measurement, this is the model that best describes the shape of the data.

\begin{figure*}[htbp]
    \centering
    \includegraphics[width=.45\linewidth]{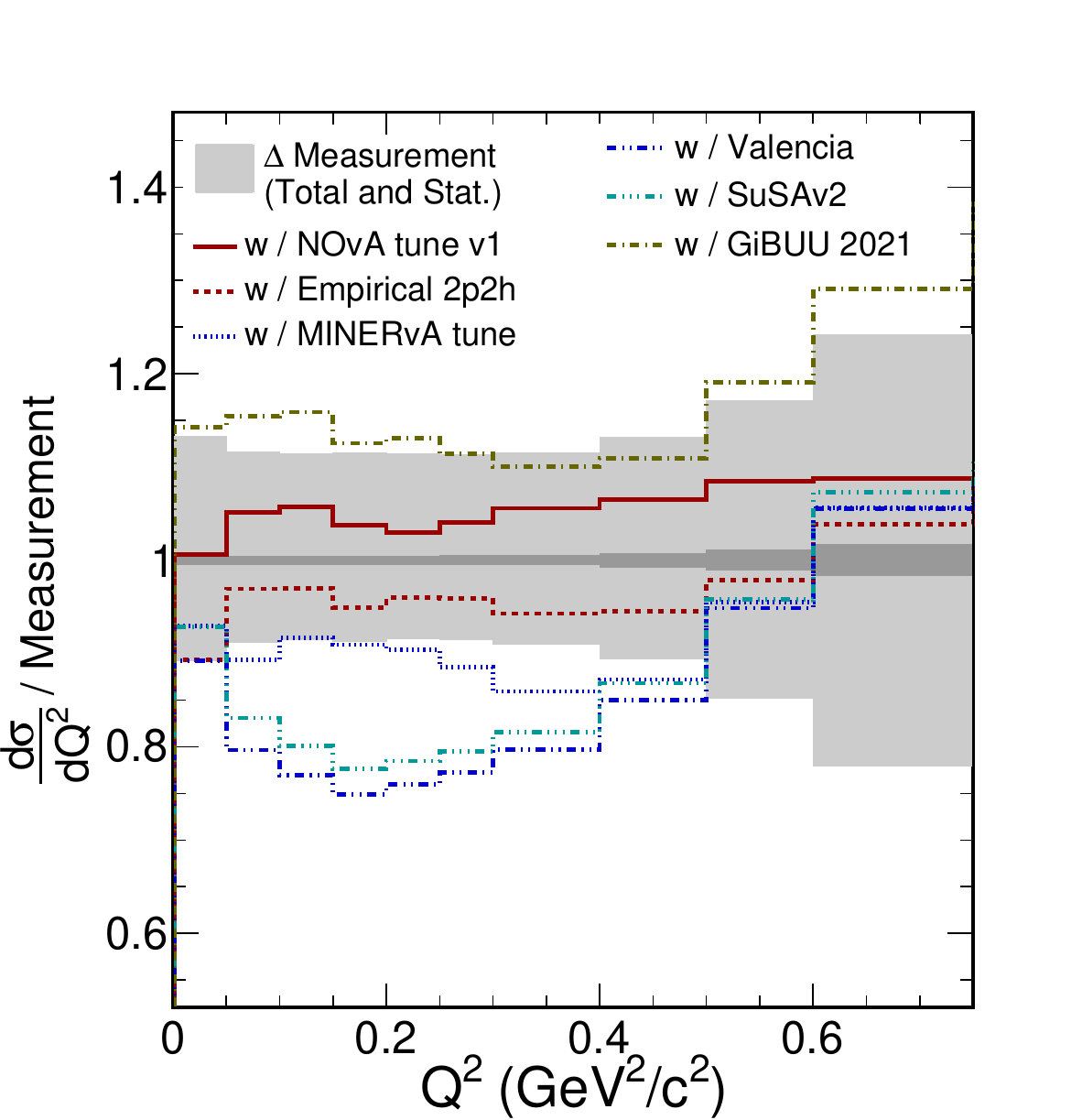}
    \includegraphics[width=.45\linewidth]{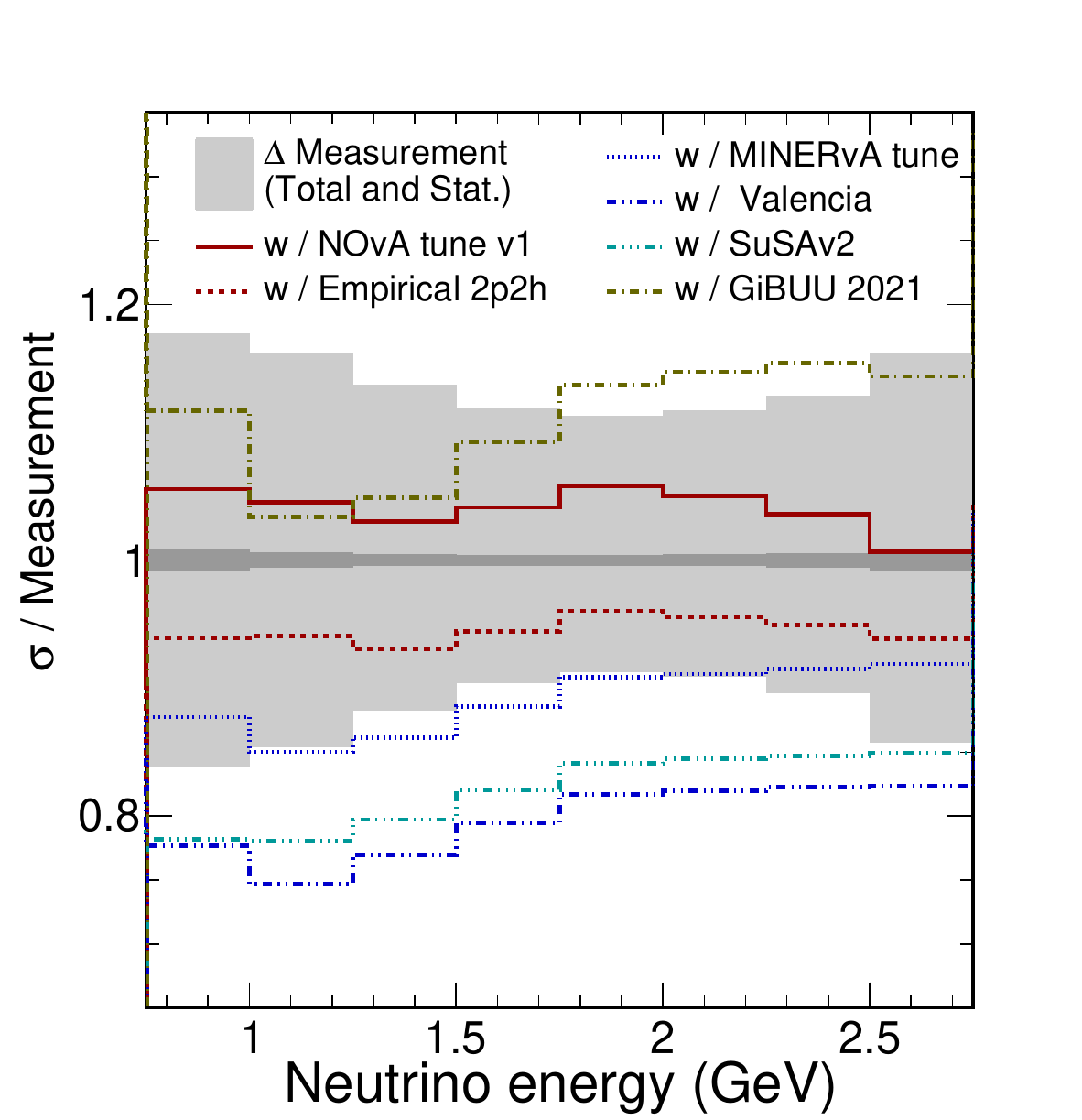}
       \includegraphics[width=.45\linewidth]{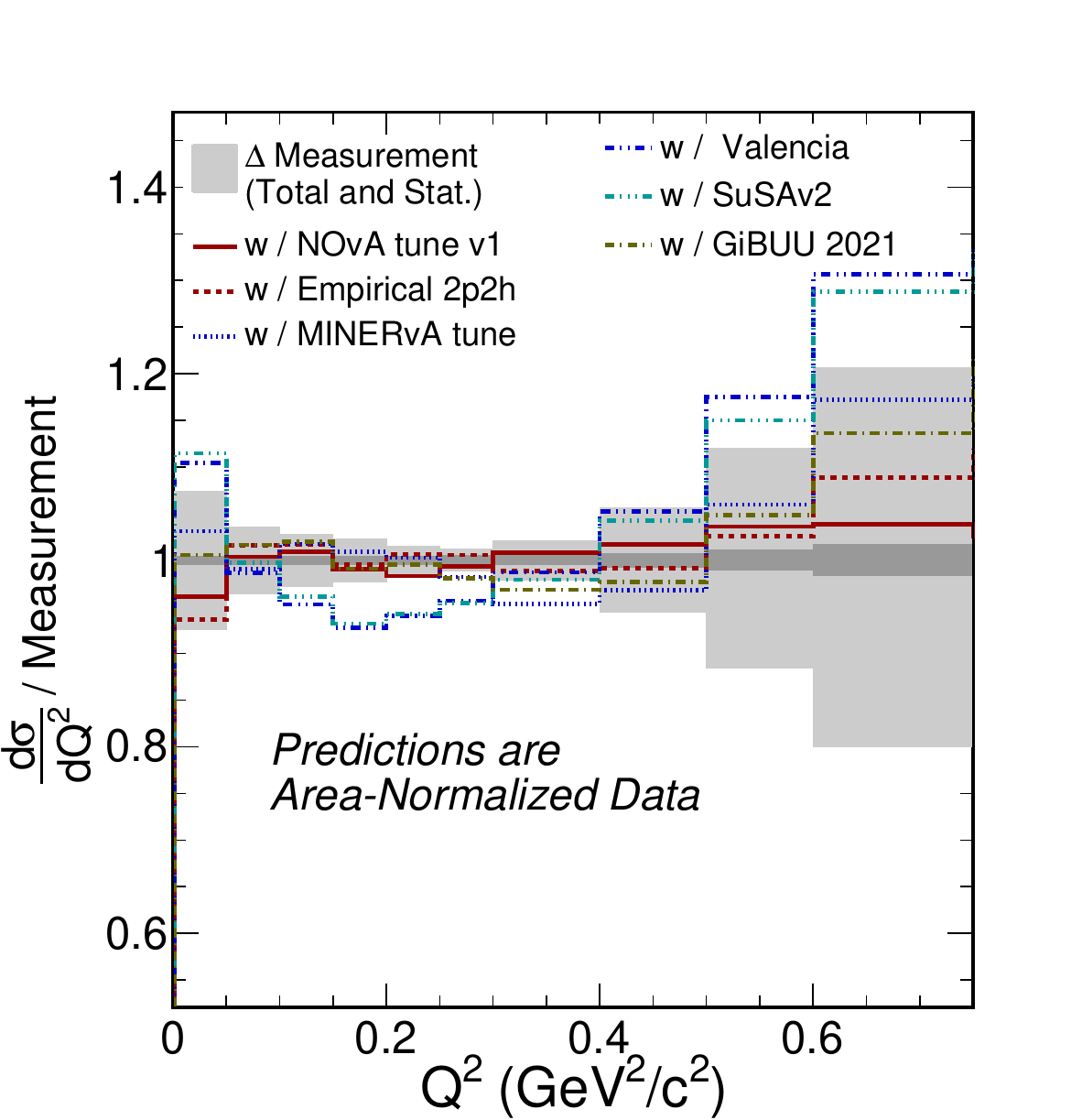}
    \includegraphics[width=.45\linewidth]{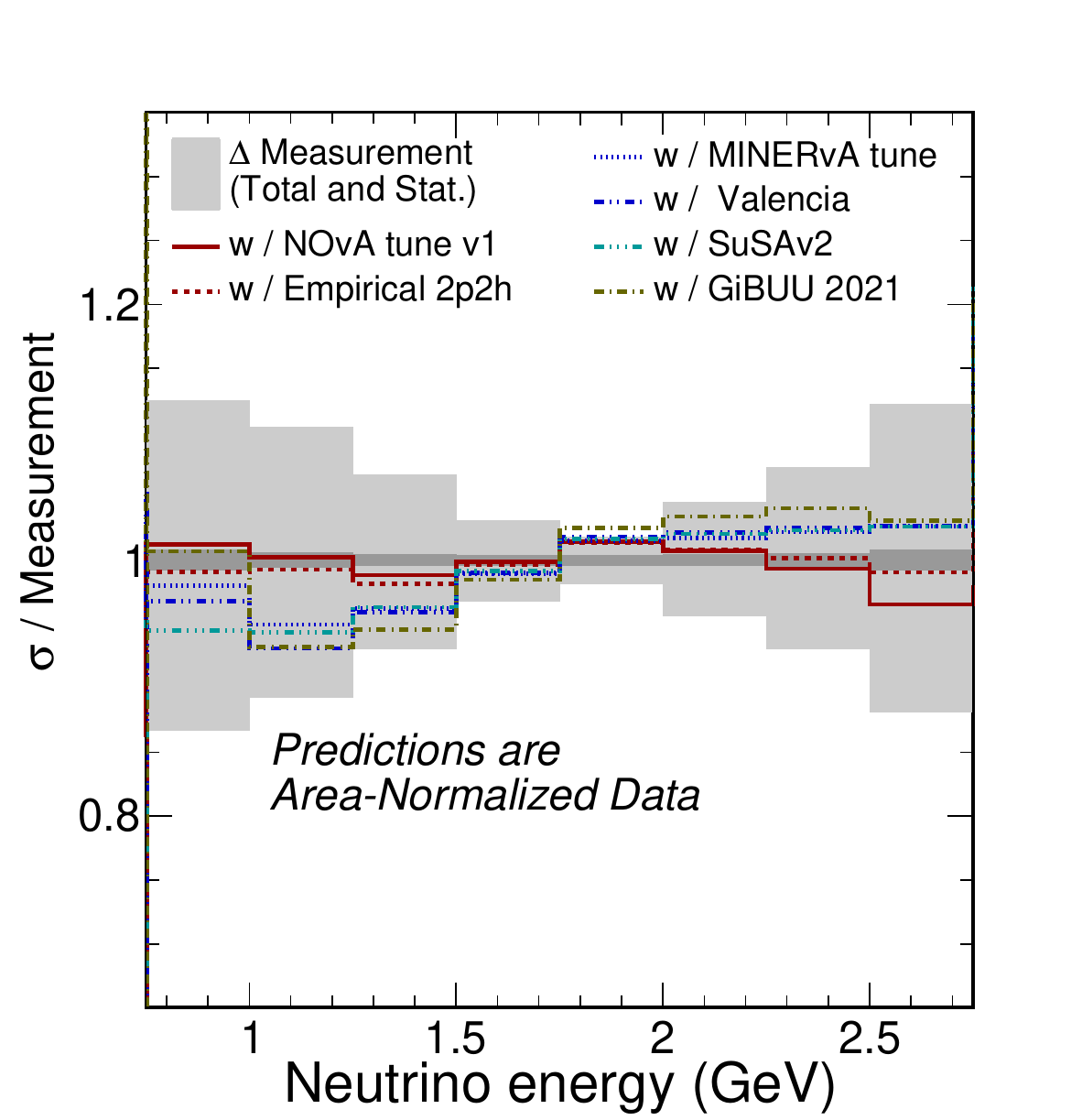}
    \caption{
      Ratios of 2p2h cross-section simulations (Empirical, \valencia, \susa, MINERvA tune (MnvTune-v1.2) and GiBUU 2021) to our single differential cross-section measurements vs.\ $Q^{2}$ (left) and vs.\ neutrino energy (right). The light (dark) gray area corresponds to the total (statistical) systematic uncertainty band. 
      Top: ratio shown with total uncertainty; bottom: ratio shown with shape-only uncertainty and models area-normalized to data.}
    \label{fig:XSecRat1D}
\end{figure*}

Figure~\ref{fig:XSecRat1D} shows the same ratios but for the single differential cross section with respect to \Qsq and the cross section as a function of the neutrino energy. 
The top plots show absolutely normalized ratios and those at the bottom show shape-only ratios.
The \valencia, MINERvA tune, and \susa models tend to underestimate both single-differential measurements; their discrepancies in the area-normalized comparisons fall within uncertainties for the neutrino energy distribution but not for \Qsq.  GiBUU largely overestimates the data, and most of the bins are close to the upper $1\sigma$ error band. The Empirical 2p2h is in better overall agreement with our measurement.

The NOvA tune v1 has the lowest $\chi^2$ values for $Q^{2}$, and is the closest to our measurement in Fig.~\ref{fig:XSecRat1D}.
The \valencia and \susa 2p2h models predict a sharp change in the \Qsq distribution at low values, which is not strongly supported by the measurement.  
The MINERvA tune model has the lowest $\chi^{2}$ in the neutrino energy comparison, despite sitting at the edges of the 1$\sigma$ band in Figure~\ref{fig:XSecRat1D}. This shows the importance of accounting for the bin-to-bin correlations when comparing models to the data. 
The theory-based models (\valencia, \susa and GiBUU 2021) predict an increase in the cross section with \Enu, whereas the measurement shows almost no neutrino energy dependence.  

\begin{widetext}
\begin{table*}[!htbp]
\begin{center}
\caption{Summary of the $\chi^2$ calculation between the data and simulations using different 2p2h models. The values in parentheses are the shape-only comparisons.}
\begin{tabular}{c|c|c|c}
\multirow{2}{*}{2p2h implementation} & $\frac{d^2\sigma}{d\cosmu dT_\mu}$ $\chi^2$& $\sigma(\Enu)$ $\chi^2$& $\frac{d\sigma}{d\Qsq}$ $\chi^2$\\
 & NDF: 115 (114) & NDF: 8 (7) & NDF: 10 (9) \\ 
\hline
NOvA tune v1     & 197 (178) & 7.5 (6.3) & 24.2 (20.0) \\
Empirical     & 190 (209) & 4.5 (4.5) & 20.8 (19.4) \\
\susa       & 499 (698) & 4.0 (1.4) & 41.6 (68.1) \\
MINERvA tune  & 330 (386) & 2.3 (2.6) & 51.1 (63.2) \\
\valencia      & 510 (756) & 6.1 (3.1) & 41.1 (64.9) \\
GiBUU         & 563 (501)     & 8.7 (7.8) & 43.1 (27.5) \\
\end{tabular}
\label{tab:chi2}
\end{center}
\end{table*}
\end{widetext}

\section{Conclusions}
\label{sec:conclusions}
This paper presents measurements of low hadronic energy \numu CC cross sections in the NOvA Near Detector that have an enhanced fraction of 2p2h and QE interactions. The analysis is based on the same simulation and reconstruction used in \cite{bib:NOvANuMuCCInc}, and was designed to have 
minimal dependence on the cross-section model. The maximum energy for final-state hadrons are 250 MeV for protons and 175 MeV for pions.  The main results are double-differential cross sections measured in 115 muon kinematics bins. Additionally, we present the single-differential cross sections with respect to the four-momentum transfer squared, and as a function of the neutrino energy. These results typically have 12\% systematic uncertainties, dominated by flux uncertainties, with almost negligible statistical uncertainties. The tabulated results can be found in Appendix \ref{app:ResultsTable} and in electronic format at the NOvA data release page \href{https://novaexperiment.fnal.gov/data-releases/}{}~\footnote{\href{https://novaexperiment.fnal.gov/data-releases/}{https://novaexperiment.fnal.gov/data-releases/}}. Future analyses will involve improved simulation and charged-pion reconstruction, enabling additional inclusive and exclusive measurements, for both neutrinos and antineutrinos.

We test 2p2h models by calculating their level of agreement with our measurements. We extend this test to include shape-only comparisons. The channel explored in this paper has an enhanced 2p2h component, expected to be more than 35\% of the total selected events in many bins, particularly in the forward-going muon region. Four 2p2h models are implemented in our simulation by substituting them in the NOvA tune v1: the \valencia, Empirical, MINERvA MnvTune-v1.2, and \susa models. We also test a GiBUU 2021 prediction (default model configuration) that incorporates a more theory-driven physics approach for interactions and particle propagation in the nucleus. None of the models accurately reproduces the measurements reported in this paper, with varying degrees of discrepancy in various regions of muon kinematic phase space, but especially in the forward-going region. The double-differential cross-section measurements presented in this paper indicate where the models need improvement and should serve as a constraint in improving neutrino--nucleus scattering models.    

\section{Acknowledgements}
This document was prepared by the NOvA collaboration using the resources of the Fermi National Accelerator Laboratory (Fermilab), a U.S. Department of Energy, Office of Science, HEP User Facility. Fermilab is managed by Fermi Research Alliance, LLC (FRA), acting under Contract No. DE-AC02-07CH11359. This work was supported by the U.S. Department of Energy; the U.S. National Science Foundation; the Department of Science and Technology, India; the European Research Council; the MSMT CR, GA UK, Czech Republic; the RAS, MSHE, and RFBR, Russia; CNPq and FAPEG, Brazil; UKRI, STFC and the Royal Society, United Kingdom; and the state and University of Minnesota.  We are grateful for the contributions of the staff of the University of Minnesota at the Ash River Laboratory, and of Fermilab. For the purpose of open access, the author has applied a Creative Commons Attribution (CC BY) license to any Author Accepted Manuscript version arising.

\FloatBarrier
\bibliography{bibs}
\bibliographystyle{apsrev4-1}

\appendix

\onecolumngrid

\section{Results in Table Format}
\label{app:ResultsTable}
 \sisetup{round-mode=places, round-precision=2}

\begin{center}
\begin{longtable}{c c c c c}
\caption{$\dd$ results table $\left(\dfrac{\mathrm{cm}^2}{\mathrm{GeV~ nucleon}} \times 10^{-39} \right)$}
\label{tab:xsecmukin}\\
\hline
\multirow{2}{*}{\bfseries $\cosmu$ range} & \multirow{2}{*}{\bfseries $T_\mu$ range (GeV)} &   \multirow{2}{*}{\bfseries Cross section} & \multirow{2}{*}{\bfseries Total Error} & \multirow{2}{*}{\bfseries Stat. Error} \\[2ex]
\hline
\endfirsthead

\caption{$\dd$ results table $\left(\dfrac{\mathrm{cm}^2}{\mathrm{GeV~ nucleon}} \times 10^{-39} \right)$}\\
\hline
\multirow{2}{*}{\bfseries $\cosmu$ range} & \multirow{2}{*}{\bfseries $T_\mu$ range (GeV)} &   \multirow{2}{*}{\bfseries Cross section} & \multirow{2}{*}{\bfseries Total Error} & \multirow{2}{*}{\bfseries Stat. Error} \\[2ex]
\hline
\endhead

\hline
\endfoot

\hline
\endlastfoot

\hline 
$[0.50,0.68)$  &  $[0.5,0.6)$  &  1.732  &  0.256  &  0.038  \\
$[0.50,0.68)$  &  $[0.6,0.7)$  &  1.597  &  0.218  &  0.047  \\
$[0.68,0.74)$  &  $[0.5,0.6)$  &  2.625  &  0.308  &  0.064  \\
$[0.68,0.74)$  &  $[0.6,0.7)$  &  2.801  &  0.367  &  0.079  \\
$[0.74,0.80)$  &  $[0.5,0.6)$  &  3.111  &  0.367  &  0.067  \\
$[0.74,0.80)$  &  $[0.6,0.7)$  &  3.662  &  0.448  &  0.080  \\
$[0.74,0.80)$  &  $[0.7,0.8)$  &  3.917  &  0.496  &  0.098  \\
$[0.74,0.80)$  &  $[0.8,0.9)$  &  3.769  &  0.552  &  0.117  \\
$[0.80,0.85)$  &  $[0.5,0.6)$  &  3.600  &  0.444  &  0.076  \\
$[0.80,0.85)$  &  $[0.6,0.7)$  &  4.567  &  0.539  &  0.097  \\
$[0.80,0.85)$  &  $[0.7,0.8)$  &  5.491  &  0.674  &  0.113  \\
$[0.80,0.85)$  &  $[0.8,0.9)$  &  5.941  &  0.731  &  0.125  \\
$[0.80,0.85)$  &  $[0.9,1.0)$  &  6.209  &  0.795  &  0.150  \\
$[0.80,0.85)$  &  $[1.0,1.1)$  &  5.867  &  0.820  &  0.178  \\
$[0.85,0.88)$  &  $[0.5,0.6)$  &  3.605  &  0.466  &  0.084  \\
$[0.85,0.88)$  &  $[0.6,0.7)$  &  4.887  &  0.604  &  0.104  \\
$[0.85,0.88)$  &  $[0.7,0.8)$  &  6.443  &  0.759  &  0.136  \\
$[0.85,0.88)$  &  $[0.8,0.9)$  &  8.071  &  0.898  &  0.158  \\
$[0.85,0.88)$  &  $[0.9,1.0)$  &  9.202  &  1.048  &  0.190  \\
$[0.85,0.88)$  &  $[1.0,1.1)$  &  9.303  &  1.091  &  0.212  \\
$[0.85,0.88)$  &  $[1.1,1.2)$  &  8.798  &  1.130  &  0.240  \\
$[0.88,0.91)$  &  $[0.5,0.6)$  &  3.387  &  0.446  &  0.083  \\
$[0.88,0.91)$  &  $[0.6,0.7)$  &  4.843  &  0.615  &  0.100  \\
$[0.88,0.91)$  &  $[0.7,0.8)$  &  6.692  &  0.767  &  0.126  \\
$[0.88,0.91)$  &  $[0.8,0.9)$  &  9.152  &  1.011  &  0.155  \\
$[0.88,0.91)$  &  $[0.9,1.0)$  &  11.333  &  1.223  &  0.188  \\
$[0.88,0.91)$  &  $[1.0,1.1)$  &  12.774  &  1.497  &  0.211  \\
$[0.88,0.91)$  &  $[1.1,1.2)$  &  13.884  &  1.543  &  0.241  \\
$[0.88,0.91)$  &  $[1.2,1.3)$  &  13.517  &  1.615  &  0.269  \\
$[0.91,0.94)$  &  $[0.5,0.6)$  &  3.141  &  0.406  &  0.089  \\
$[0.91,0.94)$  &  $[0.6,0.7)$  &  4.555  &  0.565  &  0.096  \\
$[0.91,0.94)$  &  $[0.7,0.8)$  &  6.666  &  0.804  &  0.118  \\
$[0.91,0.94)$  &  $[0.8,0.9)$  &  9.461  &  1.042  &  0.144  \\
$[0.91,0.94)$  &  $[0.9,1.0)$  &  12.895  &  1.439  &  0.182  \\
$[0.91,0.94)$  &  $[1.0,1.1)$  &  15.980  &  1.703  &  0.209  \\
$[0.91,0.94)$  &  $[1.1,1.2)$  &  19.142  &  1.929  &  0.237  \\
$[0.91,0.94)$  &  $[1.2,1.3)$  &  21.072  &  2.221  &  0.255  \\
$[0.91,0.94)$  &  $[1.3,1.4)$  &  22.081  &  2.379  &  0.285  \\
$[0.91,0.94)$  &  $[1.4,1.5)$  &  21.360  &  2.399  &  0.312  \\
$[0.91,0.94)$  &  $[1.5,1.6)$  &  18.533  &  2.301  &  0.315  \\
$[0.91,0.94)$  &  $[1.6,1.7)$  &  14.890  &  2.199  &  0.314  \\
$[0.94,0.96)$  &  $[0.5,0.6)$  &  2.794  &  0.339  &  0.081  \\
$[0.94,0.96)$  &  $[0.6,0.7)$  &  3.979  &  0.501  &  0.097  \\
$[0.94,0.96)$  &  $[0.7,0.8)$  &  6.161  &  0.777  &  0.122  \\
$[0.94,0.96)$  &  $[0.8,0.9)$  &  9.278  &  1.106  &  0.158  \\
$[0.94,0.96)$  &  $[0.9,1.0)$  &  12.964  &  1.490  &  0.186  \\
$[0.94,0.96)$  &  $[1.0,1.1)$  &  17.614  &  1.887  &  0.228  \\
$[0.94,0.96)$  &  $[1.1,1.2)$  &  22.795  &  2.483  &  0.259  \\
$[0.94,0.96)$  &  $[1.2,1.3)$  &  27.565  &  2.940  &  0.294  \\
$[0.94,0.96)$  &  $[1.3,1.4)$  &  31.144  &  3.202  &  0.327  \\
$[0.94,0.96)$  &  $[1.4,1.5)$  &  33.805  &  3.471  &  0.353  \\
$[0.94,0.96)$  &  $[1.5,1.6)$  &  33.362  &  3.444  &  0.374  \\
$[0.94,0.96)$  &  $[1.6,1.7)$  &  31.222  &  3.471  &  0.378  \\
$[0.94,0.96)$  &  $[1.7,1.8)$  &  27.041  &  3.258  &  0.375  \\
$[0.94,0.96)$  &  $[1.8,1.9)$  &  22.331  &  2.897  &  0.358  \\
$[0.94,0.96)$  &  $[1.9,2.0)$  &  16.750  &  2.445  &  0.322  \\
$[0.96,0.98)$  &  $[0.5,0.6)$  &  2.222  &  0.293  &  0.073  \\
$[0.96,0.98)$  &  $[0.6,0.7)$  &  3.458  &  0.499  &  0.090  \\
$[0.96,0.98)$  &  $[0.7,0.8)$  &  5.068  &  0.639  &  0.106  \\
$[0.96,0.98)$  &  $[0.8,0.9)$  &  7.815  &  1.043  &  0.134  \\
$[0.96,0.98)$  &  $[0.9,1.0)$  &  11.940  &  1.396  &  0.169  \\
$[0.96,0.98)$  &  $[1.0,1.1)$  &  16.898  &  1.932  &  0.202  \\
$[0.96,0.98)$  &  $[1.1,1.2)$  &  22.735  &  2.555  &  0.236  \\
$[0.96,0.98)$  &  $[1.2,1.3)$  &  29.187  &  3.056  &  0.267  \\
$[0.96,0.98)$  &  $[1.3,1.4)$  &  35.507  &  3.694  &  0.293  \\
$[0.96,0.98)$  &  $[1.4,1.5)$  &  41.157  &  4.221  &  0.321  \\
$[0.96,0.98)$  &  $[1.5,1.6)$  &  45.346  &  4.564  &  0.348  \\
$[0.96,0.98)$  &  $[1.6,1.7)$  &  46.023  &  4.554  &  0.352  \\
$[0.96,0.98)$  &  $[1.7,1.8)$  &  45.279  &  4.607  &  0.360  \\
$[0.96,0.98)$  &  $[1.8,1.9)$  &  42.685  &  4.462  &  0.364  \\
$[0.96,0.98)$  &  $[1.9,2.0)$  &  37.419  &  4.187  &  0.350  \\
$[0.96,0.98)$  &  $[2.0,2.1)$  &  31.463  &  3.859  &  0.334  \\
$[0.96,0.98)$  &  $[2.1,2.2)$  &  25.543  &  3.353  &  0.314  \\
$[0.96,0.98)$  &  $[2.2,2.3)$  &  20.728  &  2.916  &  0.316  \\
$[0.96,0.98)$  &  $[2.3,2.4)$  &  16.254  &  2.456  &  0.293  \\
$[0.98,0.99)$  &  $[0.5,0.6)$  &  1.649  &  0.240  &  0.072  \\
$[0.98,0.99)$  &  $[0.6,0.7)$  &  2.639  &  0.386  &  0.095  \\
$[0.98,0.99)$  &  $[0.7,0.8)$  &  3.933  &  0.527  &  0.116  \\
$[0.98,0.99)$  &  $[0.8,0.9)$  &  6.032  &  0.850  &  0.147  \\
$[0.98,0.99)$  &  $[0.9,1.0)$  &  9.522  &  1.305  &  0.188  \\
$[0.98,0.99)$  &  $[1.0,1.1)$  &  14.184  &  1.733  &  0.229  \\
$[0.98,0.99)$  &  $[1.1,1.2)$  &  19.255  &  2.224  &  0.262  \\
$[0.98,0.99)$  &  $[1.2,1.3)$  &  25.622  &  2.812  &  0.297  \\
$[0.98,0.99)$  &  $[1.3,1.4)$  &  32.224  &  3.583  &  0.331  \\
$[0.98,0.99)$  &  $[1.4,1.5)$  &  39.712  &  4.288  &  0.373  \\
$[0.98,0.99)$  &  $[1.5,1.6)$  &  44.884  &  4.734  &  0.402  \\
$[0.98,0.99)$  &  $[1.6,1.7)$  &  49.818  &  5.257  &  0.415  \\
$[0.98,0.99)$  &  $[1.7,1.8)$  &  52.559  &  5.274  &  0.426  \\
$[0.98,0.99)$  &  $[1.8,1.9)$  &  52.833  &  5.260  &  0.449  \\
$[0.98,0.99)$  &  $[1.9,2.0)$  &  50.114  &  5.084  &  0.443  \\
$[0.98,0.99)$  &  $[2.0,2.1)$  &  45.487  &  4.922  &  0.426  \\
$[0.98,0.99)$  &  $[2.1,2.2)$  &  40.050  &  5.125  &  0.416  \\
$[0.98,0.99)$  &  $[2.2,2.3)$  &  35.322  &  4.247  &  0.410  \\
$[0.98,0.99)$  &  $[2.3,2.4)$  &  30.891  &  3.817  &  0.419  \\
$[0.98,0.99)$  &  $[2.4,2.5)$  &  26.400  &  3.390  &  0.410  \\
$[0.99,1.00)$  &  $[0.5,0.6)$  &  1.261  &  0.212  &  0.082  \\
$[0.99,1.00)$  &  $[0.6,0.7)$  &  1.913  &  0.302  &  0.094  \\
$[0.99,1.00)$  &  $[0.7,0.8)$  &  2.799  &  0.455  &  0.109  \\
$[0.99,1.00)$  &  $[0.8,0.9)$  &  4.153  &  0.572  &  0.133  \\
$[0.99,1.00)$  &  $[0.9,1.0)$  &  6.359  &  0.850  &  0.161  \\
$[0.99,1.00)$  &  $[1.0,1.1)$  &  9.693  &  1.293  &  0.191  \\
$[0.99,1.00)$  &  $[1.1,1.2)$  &  13.718  &  1.751  &  0.227  \\
$[0.99,1.00)$  &  $[1.2,1.3)$  &  18.407  &  2.270  &  0.257  \\
$[0.99,1.00)$  &  $[1.3,1.4)$  &  23.670  &  2.862  &  0.285  \\
$[0.99,1.00)$  &  $[1.4,1.5)$  &  29.333  &  3.425  &  0.315  \\
$[0.99,1.00)$  &  $[1.5,1.6)$  &  34.279  &  3.942  &  0.331  \\
$[0.99,1.00)$  &  $[1.6,1.7)$  &  38.625  &  4.446  &  0.346  \\
$[0.99,1.00)$  &  $[1.7,1.8)$  &  41.688  &  4.706  &  0.359  \\
$[0.99,1.00)$  &  $[1.8,1.9)$  &  43.041  &  4.752  &  0.368  \\
$[0.99,1.00)$  &  $[1.9,2.0)$  &  42.559  &  4.809  &  0.376  \\
$[0.99,1.00)$  &  $[2.0,2.1)$  &  41.184  &  4.746  &  0.381  \\
$[0.99,1.00)$  &  $[2.1,2.2)$  &  38.630  &  4.502  &  0.374  \\
$[0.99,1.00)$  &  $[2.2,2.3)$  &  35.097  &  4.173  &  0.363  \\
$[0.99,1.00)$  &  $[2.3,2.4)$  &  30.622  &  3.679  &  0.360  \\
$[0.99,1.00)$  &  $[2.4,2.5)$  &  27.035  &  3.418  &  0.344  \\
\hline
\end{longtable}
\end{center}

\begin{table}
\begin{center}
\caption{$\sigma(\Enu)$ results table $\left(\dfrac{\mathrm{cm}^2}{\mathrm{nucleon}} \times 10^{-39} \right)$} 
\begin{tabular}{c c c c}
\hline
\multirow{2}{*}{\bfseries $E_\nu$ range (GeV)}&  \multirow{2}{*}{ \bfseries Cross section }& \multirow{2}{*}{\bfseries Total Error} & \multirow{2}{*}{\bfseries Stat. Error} \\[2ex] 
\hline 
$[0.75,1.00)$  &  2.844  &  0.482  &  0.022  \\
$[1.00,1.25)$  &  4.683  &  0.722  &  0.026  \\
$[1.25,1.50)$  &  5.200  &  0.662  &  0.023  \\
$[1.50,1.75)$  &  5.130  &  0.553  &  0.020  \\
$[1.75,2.00)$  &  5.114  &  0.516  &  0.020  \\
$[2.00,2.25)$  &  5.134  &  0.538  &  0.022  \\
$[2.25,2.50)$  &  5.052  &  0.591  &  0.027  \\
$[2.50,2.75)$  &  4.746  &  0.726  &  0.037  \\
\hline
\end{tabular}
\label{tab:xsecenu}
\end{center}
\end{table}

\newpage
\begin{table}
\begin{center}
\caption{$\sd$ results table $\left(\dfrac{\mathrm{cm}^2}{\mathrm{GeV^{2}~ nucleon}} \times 10^{-39} \right)$} 
\begin{tabular}{c c c c}
\hline
\multirow{2}{*}{\bfseries $Q^{2}$ range ($\mathrm{GeV^2}$)}&   \multirow{2}{*}{\bfseries Cross section} & \multirow{2}{*}{\bfseries Total Error} & \multirow{2}{*}{\bfseries Stat. Error} \\ [2ex]
\hline 
$[0.00,0.05)$  &  6.600  &  0.801  &  0.029  \\
$[0.05,0.10)$  &  9.141  &  0.945  &  0.037  \\
$[0.10,0.15)$  &  9.622  &  0.973  &  0.040  \\
$[0.15,0.20)$  &  9.483  &  0.969  &  0.041  \\
$[0.20,0.25)$  &  8.635  &  0.869  &  0.039  \\
$[0.25,0.30)$  &  7.598  &  0.763  &  0.036  \\
$[0.30,0.40)$  &  5.944  &  0.616  &  0.030  \\
$[0.40,0.50)$  &  3.725  &  0.446  &  0.027  \\
$[0.50,0.60)$  &  1.735  &  0.279  &  0.019  \\
$[0.60,0.75)$  &  0.629  &  0.146  &  0.010  \\
\hline
\end{tabular}
\label{tab:xsecq2}
\end{center}
\end{table}

\end{document}